\documentclass[preprint]{elsarticle}

\usepackage[utf8]{inputenc}
\usepackage[fleqn, cmex10]{amsmath}
\usepackage{mathtools}
\usepackage{amssymb}
\usepackage{color}
\usepackage{graphicx}
\usepackage{algpseudocode}
\usepackage{amsthm}
\usepackage{multicol}

\newtheorem*{myTh}{Theorem}

\usepackage{hyperref}
\hypersetup{
backref=true,
pagebackref=true,
hyperindex=true,
colorlinks=true,
breaklinks=true,
urlcolor= blue,
linkcolor= blue,
bookmarks=true,
bookmarksopen=true,
pdftitle={Journal version MultiSSSA},
pdfauthor={Isaac Yoann},
pdfsubject={Sparse structured approximation on over-complete dictionary}
}
\usepackage[all]{hypcap}

\newcommand{\argmin}{\operatornamewithlimits{argmin}}
\newcommand{\argmax}{\operatornamewithlimits{argmax}}
\newcommand{\softth}{\operatornamewithlimits{SoftThreshold}}
\newcommand{\etcDots}{\operatornamewithlimits{\dots}}

\usepackage{tikz}

\tikzstyle{block} = [draw, fill=blue!20, rectangle, 
    minimum height=3em, minimum width=6em, font=\small]
\tikzstyle{image} = [rectangle, 
    minimum height=18em, minimum width=18em]
\tikzstyle{input} = [coordinate]
\tikzstyle{output} = [coordinate]
\tikzstyle{pinstyle} = [pin edge={to-,thin,black}]
\tikzstyle{every node}=[font=\small]

\def\XX{{Multi-SSSA}}

\def\muu{multi-dimensional}
\def\mou{mono-dimensional}

\usepackage[caption=false,font=footnotesize]{subfig}

\begin{document}


\begin{frontmatter}


\title{Multi-dimensional signal approximation with sparse structured priors using split Bregman iterations \tnoteref{t1}}

\author[cea,lri]{Y.~Isaac}
\ead{isaac.yoann@gmail.com}

\author[mensia]{Q.~Barth{\'e}lemy}
\ead{q.barthelemy@gmail.com}

\author[cea]{C.~Gouy-Pailler}
\ead{cedric.gouy-pailler@cea.fr}

\author[lri]{M.~Sebag}
\ead{michele.sebag@lri.fr}

\author[dauphine]{J.~Atif}
\ead{jamal.atif@dauphine.fr}

\address[cea]{CEA, LIST, 91191 Gif-sur-Yvette Cedex, France}
\address[mensia]{Mensia Technologies S.A., 75015 Paris, France}
\address[lri]{Universit{\'e} Paris-Sud, TAO, CNRS $-$ INRIA $-$ LRI, 91405 Orsay, France}
\address[dauphine]{PSL, Universit{\'e} Paris-Dauphine, LAMSADE, CNRS, UMR 7243, 75775 Paris, France}

\tnotetext[t1]{The work presented in this paper has been partially funded by DIGITEO under the grant 2011-053D.}
        
\arraycolsep=3pt

\begin{abstract}
This paper addresses the structurally-constrained sparse decomposition of multi-dimensional signals onto overcomplete families of vectors,
called dictionaries. The contribution of the paper is threefold.
Firstly, a generic spatio-temporal regularization term is designed and used together with the standard $\ell_1$ regularization term
to enforce a sparse decomposition preserving the spatio-temporal structure of the signal.
Secondly, an optimization algorithm based on the split Bregman approach is proposed to handle the associated optimization problem,
and its convergence is analyzed. Our well-founded approach yields same accuracy as the other algorithms at the state-of-the-art,
with significant gains in terms of convergence speed.
Thirdly, the empirical validation of the approach on artificial and real-world problems demonstrates the generality and effectiveness
of the method. On artificial problems, the proposed regularization subsumes the Total Variation minimization and recovers the expected
decomposition. On the real-world problem of electro-encephalography brainwave decomposition, the approach outperforms similar approaches in terms of P300 evoked potentials detection, using structured spatial priors to guide the decomposition.
\end{abstract}

\begin{keyword}
Structured sparsity, overcomplete representations, analysis prior, split Bregman, fused-LASSO, EEG denoising.
\end{keyword}

\end{frontmatter}


\def\SSA{Multi-dimensional sparse signal approximation}
\def\ssa{Multi-SSA}

\section{Introduction}
\label{sec:intro}

In the last two decades, dictionary-based representations have been applied with success on a number of tasks, e.g. robust transmission with compressed sensing~\cite{donoho2006compressed}, image restoration~\cite{mairal2008sparse}, blind source separation~\cite{lee1999blind} or classification~\cite{wright2009robust} to name a few. Dictionary-based representations proceed by approximating a signal with a linear combination of elements, referred to as dictionary atoms, where the dictionary is either given based on the domain knowledge, or learned from a signal database~\cite{tosic2011dictionary}.
\\
The state-of-the-art mostly considers overcomplete dictionaries; in such cases the signal decomposition is not unique, requiring to select a decomposition
with specific constraints or properties. One such property is the {\em sparsity}, where the signal decomposition involves but a few atoms; the trade-off between the number of such atoms and the approximation error is controlled via a weighted penalization term. Despite their good properties, sparse decompositions are sensitive w.r.t. the data noise~\cite{donoho2006stable}, particularly so when the dictionary atoms are highly correlated.

This paper focuses on preserving the structure of the signal through the dictionary-based decomposition, thereby expectedly decreasing its sensitivity w.r.t. noise. The proposed approach is motivated by, and illustrated on, spatio-temporal multi-dimensional signals, referred to as multi-channel signals. The decomposition of multi-channel signals involves: i) decomposing each channel into the dictionary (see \cite{rakotomamonjy2011surveying} for a survey); ii) ensuring that the structure of the multi-channel data is preserved in the multi-dimensional decomposition.

Following and extending previous work \cite{isaac2013multi}, this paper focuses on {\bf structured dictionary-based decomposition}, where the
dictionary-based representation preserves the structure of the signal, as follows.
Assuming that each one of the \mou~signals is structured (e.g. being recorded in consecutive time samples and being continuous w.r.t. time), the structured decomposition property
aims at preserving the signal structure in the dictionary-based representation (e.g. requiring that the approximation of signals in consecutive time steps is expressed on same atoms with ``close'' weights). Formally, the structured decomposition property is enforced via considering a specific regularization term besides the data fitting term (minimization of the approximation error) and the standard $\ell_1$ term (maximizing the decomposition sparsity).
\\
The main contribution of this paper is to propose an efficient optimization scheme based on the split Bregman iterations for performing this structured decomposition of \muu~signals. Implementation details including an efficient heuristic for hyper-parameters tuning is provided to accelerate the proposed algorithm.
The proposed Multi-dimensional Sparse Structured Signal Approximation (\XX) is assessed on synthetic signals for the fused-LASSO regularization obtained when the analysis term is a TV penalty. The approach is assessed in terms of i) its computational cost compared with the state-of-the-art smooth proximal gradient~\cite{chen2012efficient} approach;  ii) its ability to recover the sparse structure of the initial signals compared with standard sparsity constraints and fused LASSO ($\ell_0$, $\ell_{2, 0}$, $\ell_1$, $\ell_{2,1}$ and $\ell_1 + \ell_{2,1}$) even though the true structure of the signal is used to define the fused-LASSO regularization.
Finally, \XX\ is applied on electroencephalographic (EEG) signals to detect P300 evoked potentials. Using a data-driven prior, the structured decomposition approach is shown to effectively denoise the signal, resulting in a better classification accuracy compared to the other regularizations.

The paper is organized as follows. Section~\ref{sec:prob} introduces the formal background. Related work is discussed in Section~\ref{sec:rel_work}. Section \ref{sec:optim} gives an overview of the proposed \XX\ algorithm and the optimization strategy. The experimental validation of the approach on artificial data in terms of computational cost and structure recovery is presented and discussed respectively in Sections~\ref{sec:speed} and \ref{sec:exp_eval_synth}. Section~\ref{sec:EEG_application} presents the denoising application on EEG signals. The paper concludes with a general discussion and points out some perspectives in Section~\ref{sec:discussion}.

\paragraph{\bf Notations.}
In the following, the $j$-th column of a matrix $X$ is written $X(j)$, the $i$-th row $X^T(i)$, and the $i$-th element of $j$-th column $X(i,j)$. $I_n$ stands for the identity matrix of size $n$. The $\ell_p$ matrix norm is defined as $\|X\|_p = (\sum_{i}\sum_{j}|X(i,j)|^p)^{\frac{1}{p}}$, with $p=2$ corresponding to the classical Frobenius norm, and the $\ell_{p,q}$ mixed norm is defined as $\|X\|_{p,q} = (\sum_{i} (\sum_{j}|X(i,j)|^p)^{\frac{q}{p}})^{\frac{1}{q}}$.


\section{Sparse structured decomposition problem}
\label{sec:prob}

\subsection{General problem}

Let $Y=[Y(1), \dots, Y(T)] \in \mathbb{R}^{C \times T}$ be a matrix of $T$ ordered (e.g. corresponding to consecutive samples)
$C$-dimensional signals, and $\Phi \in \mathbb{R}^{C \times N_{\Phi}}$ an overcomplete dictionary of $N_{\Phi}$ normalized atoms ($N_{\Phi} \gg C$). We consider the following linear model:
\begin{flalign}
	Y(t) ~=~& \Phi X(t) + E(t),~~t \in \{1, \dots, T\} \enspace \nonumber,
	\\
	Y ~=~& \Phi X + E \enspace,
\end{flalign}
in which $X=[X(1), \cdots, X(T)] \in \mathbb{R}^{N_{\Phi} \times T}$ is the decomposition matrix 
and $E=[E(1), \cdots, E(T)] \in \mathbb{R}^{C \times T}$ stands for a Gaussian noise matrix.\\

The sparse structured approximation problem consists in decomposing each signal $Y(t)$, $t \in \{1, \cdots, T\}$ onto the dictionary such that the joint decomposition $X$ reflects the underlying structure of $Y$. This decomposition aims at removing the Gaussian noise present in these signals and at separating theirs components into target and non-target ones (depending on the considered application). The structured decomposition problem is formalized as the minimization of the objective function:
\begin{align}
	\min\limits_{X \in \mathbb{R}^{N_{\Phi} \times T}} \|Y-\Phi X\|_2^2 + \lambda_1 \|{X}\|_1 + \lambda_2 \|{X} P\|_1 \,\enspace ,
	\label{eqn:basicProblem}
\end{align}
with $\lambda_1$, $\lambda_2$ the regularization coefficients and $P \in \mathbb{R}^{T \times N_{P}}$ a matrix encoding the prior knowledge about the signal structure. The use of the  $\|XP\|_1$ regularization term can be interpreted in terms of sparse analysis \cite{elad2007analysis}. Classically, the sparse analysis decomposition problem is formalized as follows:
\begin{align}
	\min_{X \in \mathbb{R}^{C \times T}} \|Y-X\|_2^2 + \lambda_1 \|\Psi X\|_1 \,\enspace ,
\end{align}
whereas the synthesis problem writes as $\min_X \|Y - \Phi X\|_2^2 + \lambda_1 \|X\|_1$. 
In the analysis setting, atoms in the dictionary $\Psi$ are viewed as filters, on which the projection of the decomposed signals are required to be sparse. Along this line, the $P$ matrix can be viewed as a set of linear filters, on which the projections of the decomposed signals are required to be sparse, thereby preserving the regularities given from prior knowledge about the application domain. The matrix $P$ can be learned from a set of signals~\cite{peyre2011learning, rubinstein2012k}.

The present paper considers a synthesis formulation of the problem, regularized by an analysis term. In order to illustrate the interest of such decomposition, the particular case of the \muu{} fused-LASSO is described below.

\subsection{Multi-dimensional fused-LASSO problem}

In the particular case of a piecewise constant prior, a block-wise decomposition is expected. Formally this structure encoded by coefficients is defined as follows:
\begin{align}
	\forall n \in \{1, \cdots, N_{\Phi}\},~~X^T(n) = \sum_{m=1}^{M_n} \alpha_m^n \mathbf{1}_{\kappa_m^n} \ ,
\end{align}
with $\{\kappa_m^n,~\forall m \in \{1,\cdots,M_n\}\}$ a partition of $\{1, \cdots, T\}$ corresponding to the $M_n$ blocks of the coefficients associated with the $n$-th atom of $X$, and $\{\alpha_m^n \in \mathbb{R},~\forall m \in \{1, \cdots, M_n\}\}$ the coefficients associated with the blocks.
\\
In the studied model, adding an $\ell_1$ and a \muu~TV regularization terms allows to enforce such prior. The TV term leads to a sparse decomposition gradient, thus preserving signal singularities and data edges, and supporting the detection of abrupt changes~\cite{rudin1992nonlinear}. The analysis term is then written:  $\|{X} P^{\text{TV}}\|_1 ~~=~~ \sum_{t=2}^{T}\|X(t) - X(t-1)\|_1$, where
\begin{align}
P^{\text{TV}} = \begin{pmatrix}
  -1 &    &        & \\
  1  & -1 &        & \\
     & 1  & \ddots & \\
     &    & \ddots & -1 \\ 
     &    &        & 1
 \end{pmatrix} \ \in \mathbb{R}^{T \times {T-1}} \ . \nonumber
\end{align}

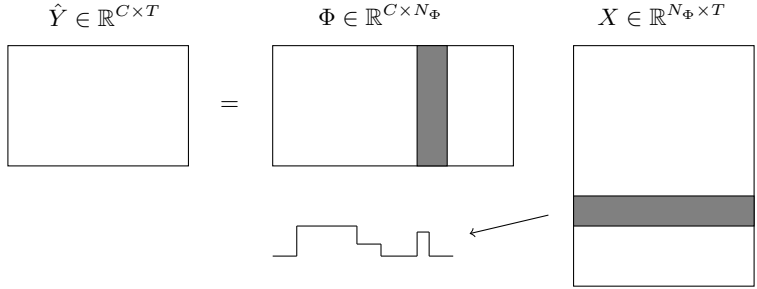
\begin{figure}[h!]
	\begin{center}
		\begin{tikzpicture}[scale=1.6]

\draw (-0.7,0) rectangle + (1.5,1);
\draw (1.3,0.5) node[left] {$=$};
\draw (1.5,0) rectangle + (2,1);
\draw (2.7,0)[fill=gray]--(2.95,0)--(2.95,1)--(2.7,1)--cycle; 
\draw (4,-1) rectangle + (1.5,2);
\draw (5.5,-0.5)[fill=gray]--(4,-0.5)--(4,-0.25)--(5.5,-0.25)--cycle;
\draw [<-] (4.6,0.9,3.8) -- (5.1,0.9,3.4);
\draw (1.5,-0.75) -- (1.7,-0.75);
\draw (1.7,-0.75) -- (1.7,-0.5);
\draw (1.7,-0.5) -- (2.2,-0.5);
\draw (2.2,-0.5) -- (2.2,-0.65);
\draw (2.2,-0.65) -- (2.4,-0.65);
\draw (2.4,-0.65) -- (2.4,-0.75);
\draw (2.4,-0.75) -- (2.7,-0.75);
\draw (2.7,-0.75) -- (2.7,-0.55);
\draw (2.7,-0.55) -- (2.8,-0.55);
\draw (2.8,-0.75) -- (2.8,-0.55);
\draw (2.8,-0.75) -- (3.0,-0.75);
\draw (0.65,1.25) node[left] {$\hat{Y} \in \mathbb{R}^{C \times T}$};
\draw (3,1.25) node[left] {$\Phi \in \mathbb{R}^{C \times N_{\Phi}}$};
\draw (5.35,1.25) node[left] {$X \in \mathbb{R}^{N_{\Phi} \times T}$};
       
		\end{tikzpicture}
	\end{center}
	\caption{Decomposition of a multi-dimensional~time-series signal $\hat{Y} = \Phi X$, with a TV regularization enforcing a block-wise structure of the coefficients X.}
	\label{fig:schem_decomp}
\end{figure}

The Figure~\ref{fig:schem_decomp} depicts an example of time-series coefficients obtained in this case, \textit{i.e.} with sparse variations, for a given atom of the dictionary (in grey). 
This combination of regularization terms is known as the fused-LASSO and has been chosen for the evaluation of the proposed algorithm on synthetic data because of its interest for various applications: frequency hopping~\cite{angelosante2010multiple}, geophysical studies~\cite{gholami2010regularization}, multi-task learning~\cite{zhou2012modeling}, trend analysis~\cite{kim2009ell_1}, covariance estimation~\cite{danaher2014joint}, analysis of association between genetic markers and traits~\cite{kim2009statistical} or change point detection~\cite{bleakley2011group}, among others.

If the fused-LASSO is a well-known model, it is important to notice that the introduced \XX\ is able to solve the general problem of Eq.~(\ref{eqn:basicProblem}) for any matrix $P$, with an analytic or a data-driven content.


\section{Related work}
\label{sec:rel_work}

Quite a few methods have been designed in the last decade to achieve the sparse approximation~\cite{cotter2005sparse} of \muu~signals while the mono-dimensional case has been intensively studied. Two main approaches have been considered: greedy methods trying to approximate the solution of the $\ell_0$ regularized problem~\cite{tropp2006algorithms1} and convex optimization solvers working on the $\ell_1$ relaxed problem~\cite{tropp2006algorithms2, gribonval2008atoms, rakotomamonjy2011surveying}. The concept of structured sparsity then emerged from the integration of prior knowledge encoded as regularizations into these approximation problems~\cite{huang2011learning, jenatton00377732}.
\\
The proposed \XX\ approach falls in the latter category with a regularization combining a classical sparsity term and an analysis one. Some analysis regularizations have been extensively studied like the TV one which has been introduced in the ROF model~\cite{rudin1992nonlinear, darbon2005fast} for image denoising, nonetheless in the context of dictionary-based decomposition, their study has only begun recently~\cite{selesnick2009signal, vaiter2013robust, majumdar2012synthesis}. Their interest for retrieving the underlying sparse structure of signals has been shown in~\cite{candes2011compressed}. To the best of our knowledge, the combination of these regularization terms had not been studied except for the particular case of the fused-LASSO introduced in~\cite{tibshirani2005sparsity}.

Despite the convexity of the associated minimization problem, the two $\ell_1$ non-differentiable terms make it difficult to solve (by classical gradient approaches). Various approaches have been developed for solving the fused-LASSO problem. In their seminal work, Tibshirani \textit{et al.}~\cite{tibshirani2005sparsity} transformed the problem to a quadratic one and used standard optimization tools. While this approach proved computationally feasible for small-sized problems since it relies on
increasing the dimension of the search space, it does not scale efficiently with problem size. 
Path algorithms have then been developed: Hoefling proposed in \cite{hoefling2010path} a method solving this problem in the particular case
of the fused-LASSO signal approximator ($\Phi = I_C$) and Tibshirani \textit{et al.}~\cite{tibshirani2011solution} designed a path method for the generalized LASSO problem. 
More recently, scalable approaches based on proximal sub-gradient methods \cite{liu2010efficient}, ADMM\footnote{Alternating Direction Method of Multipliers.}\cite{wahlberg2012admm} and split Bregman iterations \cite{ye2011split} have been successfully applied to the \mou~generalized fused-LASSO.
\\
Concerning the \muu~fused-LASSO, an efficient method has been proposed recently in~\cite{chen2012efficient} for multi-task regression. A proximal method~\cite{nesterov2005smooth} is applied to a smooth approximation of the fused-LASSO regularization terms in this study and can be considered for different analysis regularizations. This approach is the most comparable with the present work.

\begin{figure}[h!]
	\centering
	\includegraphics[width=0.95\linewidth]{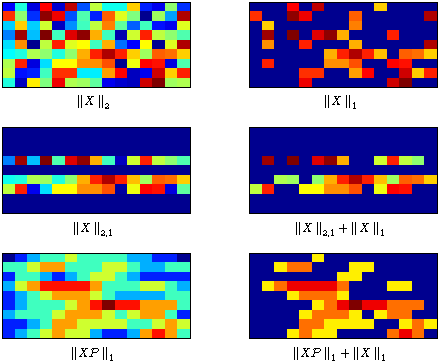}
	\label{fig:differencesBetweenRegularizations}
	\caption{Illustration of the matrix $X \in \mathbb{R}^{N_{\Phi} \times T}$ with the different regularizations: $\ell_2$ (top left), $\ell_1$ (top right), $\ell_{2,1}$ (middle left), $\ell_1+\ell_{2,1}$ (middle right), TV (bottom right) and $\ell_1+\text{TV}$ (bottom right).}
\end{figure}

To conclude this state-of-the-art, an illustration of the different regularizations is plotted in Figure~\ref{fig:differencesBetweenRegularizations}: $\ell_2$ (top left), $\ell_1$ \cite{tibshirani1996regression} (top right), $\ell_{2,1}$ \cite{yuan2006model} (middle left), $\ell_1+\ell_{2,1}$ \cite{gramfort2013time} (middle right), TV \cite{rudin1992nonlinear} (bottom right) and $\ell_1+\text{TV}$ (bottom right).

This figure also illustrates the interest of structurally constrained regularizations in various noisy situations. Many concrete applications entail extracting specific activities (signal space or target activities) from the original signals, therefore getting rid of various kind of noise, defined as spurious/non target activities. While target and non-target activities often share a few properties (statistical distribution, spatial and temporal structures), the introduction of structurally constrained regularizations aims at drastically limiting the resulting space to signals exhibiting the desired properties.


\section{Optimization strategy}
\label{sec:optim}

The split Bregman approach has been shown particularly well suited to $\ell_1$ minimization problems~\cite{goldstein2009split} because of its ability to early detect zero and non-zero coefficients, which insures it a fast convergence on these problems. Thus, this optimization scheme has been chosen to solve the studied decomposition of Eq.~(\ref{eqn:basicProblem}).

\subsection{Optimization Scheme}

The above minimization problem is written as follows,
\begin{align}
	\min\limits_{ X \in \mathbb{R}^{N_{\Phi} \times T}} &\|Y-\Phi X\|_2^2 + \lambda_1 \|X\|_1 + \lambda_2 \|XP\|_1 \nonumber\enspace .
\end{align}
To setup the optimization scheme, let us first restate it as:
\begin{align}
	\min\limits_{ X \in \mathbb{R}^{N_{\Phi} \times T} \atop A \in \mathbb{R}^{N_{\Phi} \times T}, B \in \mathbb{R}^{N_{\Phi} \times N_{P}}} &\|Y-\Phi X\|_2^2 + \lambda_1 \|A\|_1 + \lambda_2 \|B\|_1 \nonumber
	\\
	\text{s.t.~} &A = X ~~\text{and}~~ B = X P \enspace .
	\label{eqn:base}
\end{align}
This reformulation is a key step of the split Bregman method. It decouples the three terms and allows to optimize them separately within the iterations. To setup this iteration scheme, Eq.~(\ref{eqn:base}) is rewritten as an unconstrained problem:
\begin{align}
	\min\limits_{ X \in \mathbb{R}^{N_{\Phi} \times T} \atop A \in \mathbb{R}^{N_{\Phi} \times T}, B \in \mathbb{R}^{N_{\Phi} \times N_{P}}} &\|Y-\Phi X\|_2^2 + \lambda_1 \|A\|_1 + \lambda_2 \|B\|_1 \nonumber
	\\
	&~+ \frac{\mu_1}{2} \|X - A\|_2^2 + \frac{\mu_2}{2} \|X P - B\|_2^2 \nonumber \enspace .
\end{align}

\noindent Denoting $i$ the current iteration, the split Bregman scheme~\cite{goldstein2009split} is then written:
\begin{align}
	\label{eqn:primal}
	(X^{i+1}, A^{i+1}, B^{i+1}) =& \argmin\limits_{ X \in \mathbb{R}^{N_{\Phi} \times T} \atop A \in \mathbb{R}^{N_{\Phi} \times T}, B \in \mathbb{R}^{N_{\Phi} \times N_{P}}} \|Y-\Phi X\|_2^2 + \lambda_1 \|A\|_1 + \lambda_2 \|B\|_1 ~~\enspace\nonumber
	\\
	& + \frac{\mu_1}{2} \|X - A + D_{A}^i\|_2^2 + \frac{\mu_2}{2} \|X P - B + D_{B}^i\|_2^2 ~~\enspace
	\\
	D_{A}^{i+1} =& ~~~~ D_{A}^{i} + (X^{i+1} - A^{i+1})~~\enspace \nonumber
	\\
	D_{B}^{i+1} =& ~~~~ D_{B}^{i} + (X^{i+1} P - B^{i+1}) ~\nonumber \enspace .
\end{align}

This scheme is equivalent to the one obtained with the augmented Lagrangian method (or ADMM) when the constraints are linear~\cite{wu2010augmented}. Each iteration requires the solving of the primal problem before updating the dual variables.\\
Thanks to the split of the three terms, the minimization of the primal problem Eq.~(\ref{eqn:primal}) can be performed iteratively by alternatively updating
the following variables:
\begin{align}
	\label{eqn:theta_update}
	X^{i+1} &= \argmin\limits_{X \in \mathbb{R}^{N_{\Phi} \times T}} ~~~ \|Y-\Phi X\|_2^2 + \frac{\mu_1}{2} \|X - A^i + D_{A}^i\|_2^2 \nonumber 
	\\
	& ~~~~~~~~~~~~~~~~~~~ + \frac{\mu_2}{2} \|X P - B^i + D_{B}^i\|_2^2 
	\\
	\label{eqn:A_update}
	A^{i+1} &= \argmin\limits_{A \in \mathbb{R}^{N_{\Phi} \times T}} ~~~~\lambda_1 \|A\|_1 + \frac{\mu_1}{2} \|X^{i+1} - A + D_{A}^i\|_2^2 
	\\
	\label{eqn:B_update}
	B^{i+1} &= \argmin\limits_{B \in \mathbb{R}^{N_{\Phi} \times N_{P}}} ~~ \lambda_2 \|B\|_1 + \frac{\mu_2}{2} \|X^{i+1} P - B + D_{B}^i\|_2^2  \enspace .
\end{align}
Empirically, it has been noted that only few iterations of this
system are necessary for convergence~\cite{goldstein2009split}. In our
implementation, this update is only performed once at each iteration of the
global optimization algorithm.\\

\noindent Eq.~(\ref{eqn:A_update})~and~Eq.~(\ref{eqn:B_update}) can be solved with the soft-thresholding operator~\cite{combettes2005signal}:
\begin{align}
	A^{i+1} & = ~ \softth_{\frac{\lambda_1}{\mu_1}} (X^{i+1}  +  D_{A}^i),
	\\
	B^{i+1} & = ~ \softth_{\frac{\lambda_2}{\mu_2}} (X^{i+1} P  + D_{B}^i) \enspace
\end{align}
with 
\begin{align*}
	(\softth_{\lambda}(X))(i,j) = \max\left(0,\ 1-\frac{\lambda}{\left|X(i,j)\right|}\right)X(i,j).
\end{align*}
Solving Eq.~(\ref{eqn:theta_update}) requires the minimization of a convex differentiable function which can be performed via classical
optimization methods. We propose here to solve it deterministically which is an original contribution of this work. 
Let us define $H$ from Eq.~(\ref{eqn:theta_update}) such as:
\begin{align}
\label{eqn:impEqn}
	X^{i+1} = \argmin\limits_{X \in \mathbb{R}^{N_{\Phi} \times T}} H(X) \enspace .
\end{align}
Differentiating this expression with respect to $X$ yields:
\begin{align}
	\frac{d}{dX}H(X) =~~&(2\Phi^T \Phi + \mu_1 I_{N_{\Phi}})X + X (\mu_2 PP^T) \\
	& - 2\Phi^T Y  + \mu_1(D_{A}^i - A^i) + \mu_2(D_{B}^i - B^i)P^T \nonumber \enspace ,
\end{align}
The minimum $\hat{X}=X^{i+1}$ of Eq.~(\ref{eqn:theta_update}) is
obtained by solving $\frac{d}{dX}H(\hat{X}) = 0$, which is known as a Sylvester equation:
\begin{align}
	\label{eqn:syl}
	W\hat{X} + \hat{X} Z = M^i \enspace ,
\end{align}
with $W=2\Phi^T \Phi + \mu_1 I_{N_{\Phi}} \in \mathbb{R}^{N_{\Phi} \times N_{\Phi}}$, $Z=\mu_2 PP^T \in \mathbb{R}^{T \times T}$ and $M^i= 2\Phi^T Y + \mu_1 (A^i-D_{A}^i) + \mu_2 (B^i - D_{B}^i)P^T \in \mathbb{R}^{N_{\Phi} \times T}$.

In a general setting, solving efficiently a Sylvester equation can be time-consuming when the dimension of these matrices are large. A closed form solution derived from the vectorized formulation can be computed but it requires heavy calculations. One of the most used methods to deal with this issue, is the Bartels–Stewart algorithm~\cite{bartels1972solution} whose time complexity is $O(N^3_{\Phi})$.
In our particular case, the structures of the involved matrices ease this update.

Indeed, $W$ and $Z$ being real symmetric matrices, solving for $\hat{X}$ the Sylvester equation $W\hat{X}+\hat{X}Z=M^i$ is equivalent to solve for $\hat{X}'$ the following diagonal system
\begin{align}
	\label{eqn:diagSyl}
	D_w\hat{X'} + \hat{X'} D_z = M^{i}{}' \enspace ,
\end{align}
where
\begin{align}
	\label{eqn:diagonalization}
	W &= F D_w F^T,& \enspace Z &= G D_z G^T \enspace ,
	\\
	\hat{X'}&= F^T \hat{X} G ,& \enspace {M^i}{}' &= F^T M^i G \nonumber \enspace .
\end{align}
$W$ and $Z$ can be diagonalized in orthogonal bases~($F$ and $G$) which allow to rewrite the Sylvester equation as follows:
\begin{align}
	F D_w F^T\hat{X} + \hat{X} G D_z G^T=M^i \nonumber \enspace ,
\end{align}
and by applying the orthogonal properties of $F$ and $G$ the above expression is obtained.

\noindent The solution $\hat{X'}$ of the problem Eq.(\ref{eqn:diagSyl}) is then computed as follows:
\begin{align}
	\forall t \in \{1, \cdots, T\}~~\hat{X}{}'(t) = (D_w + D_z(t,t)I_{N_{\Phi}})^{-1} M^{i}{}'(t) \ , \nonumber
\end{align}
which is equivalent to
\begin{align}
	\forall n \in \{1, \cdots, N_{\Phi}\}, ~~\forall t \in \{1, \cdots, T\} \ , \nonumber 
	\\
	(D_w(n,n) + D_z(t,t))~\hat{X}{}'(n,t)=~M^{i}{}'(n,t) \ ,
\end{align}
which can be computed with
\begin{align}
	\hat{X}{}' = M^{i}{}' \oslash O \ ,
\end{align}
where $\oslash$ corresponds to an element-wise division and 
\begin{align}
	\label{eqn:computeO}
	O(n,t) = D_w(n,n) + D_z(t,t) \enspace . 
\end{align}
$\hat{X}$ is then calculated as $\hat{X} = F\hat{X}{}' G^T$.

This last update can be numerically unstable when the elements of $O$ are close to $0$. As a consequence, $\mu_1$ and $\mu_2$ should be chosen carefully to avoid numerical instabilities. In addition, some terms can be precomputed to
speed up the computations performed during each iteration. These implementation details and the choice of the penalty parameters are discussed in Section~\ref{sec:implem_penalty}, and the full algorithm is summarized in Section~\ref{subsec:sumup}.

\subsection{Convergence}
\label{subsec:conv}

Thanks to the exact solving of the primal subproblem presented in the previous section, the convergence of the above scheme can be derived by drawing inspiration from the steps described in~\cite{cai2009split}.

\begin{myTh}
\label{eqn:th_}
Assume that $\lambda_1 \geq 0$, $\lambda_2 \geq 0$, $\mu_1 > 0$ and $\mu_2 > 0$, then the following holds
\begin{align}
	\label{eqn:th1}
	\lim_{i \to \infty}  &\|Y - \Phi X^i\|_2^2 + \lambda_1 \|X^i\|_1 + \lambda_2 \|X^i P\|_1 \nonumber 
	\\
	= &\|Y - \Phi \hat{X}\|_2^2 + \lambda_1 \|\hat{X}\|_1 + \lambda_2 \|\hat{X} P\|_1
\end{align}
where $\hat{X}$ denotes the solution of our problem.\\
In addition, if our problem has a unique solution, from the convexity of $E(X) = \|Y - \Phi X\|_2^2 + \lambda_1 \|X\|_1 + \lambda_2 \|X P\|_1$ and Eq.~(\ref{eqn:th1}), we have
\begin{align}
	\lim_{i \to \infty}  X^i = \hat{X} \enspace .
\end{align}
\end{myTh}

\noindent Complete proof is available in \ref{appendix1}, demonstrating that split Bregman scheme converges to a solution of the convex problem. The uniqueness of the solution depend on the choice of $P$.

\subsection{Implementation details and penalty parameters tuning}
\label{sec:implem_penalty}

The terms $W$ and $Z$ in Eq.~(\ref{eqn:syl}) are independent of the considered iteration $i$. Their diagonalizations can then be performed only once and for all as well as the computation of $O$. These diagonalizations are
derived easily from those of $2\Phi^T \Phi$ and $PP^T$ and can be realized
off-line (and then pre-computed)
\begin{align}
	\label{eqn:computeDelta}
	2\Phi^T \Phi ~&= F \Delta_w F^T  \text{,} &PP^T &= G \Delta_z G^T 
	\\
	\label{eqn:computeDiagonal}
	D_w &= \Delta_w + \mu_1 I_{N_{\Phi}} \text{ and } &D_z &= \mu_2 \Delta_z.
\end{align}
Thus, the update in Eq.~(\ref{eqn:impEqn}) does not require heavy computations, even when the penalty parameters $\mu_1$ and $\mu_2$ change during the iterations (see below).

$Y_{\Phi} = 2F^T \Phi^T Y G$ and $P_G=P^TG$ can also be pre-computed to avoid useless calculations. Besides, one can notice that the computational cost of each iteration depends on chain multiplications of three matrices (e.g $FX^{temp} G^T$). The computation time of these products depends on the order in which the multiplications are performed. The computational costs of the three chains appearing in the previously described scheme for both multiplication orders are presented in Figure~\ref{fig:costs}.
\begin{figure}[h!]
\label{fig:costs}
	\begin{center}
		\begin{tabular}{|c|c|c|}
			\hline
			Chain & Cost LR & Cost RL\\
			\hline
			$F^T(D_A^i-A^{i})G$ & $N_{\Phi}T(N_{\Phi}+ T)$ & $N_{\Phi}T(N_{\Phi}+ T)$ \\
			\hline
			$FX^{i} G^T$ & $N_{\Phi}T(N_{\Phi}+ T)$ & $N_{\Phi}T(N_{\Phi}+ T)$ \\
			\hline
			$F^T(D_B^i-B^{i})P_G$ & $N_{\Phi} N_{P}(T+N_{\Phi})$ & $N_{\Phi} T(N_{\Phi} + N_{P})$ \\
			\hline
		\end{tabular}
		\caption{Computation costs of multiplication chains. LR: $ABC$ computed with $(AB)C$; RL: $ABC$ computed with $A(BC)$.}
	\end{center}
\end{figure}

\noindent The costs of the first two chains do not depend on the order of computation but the last one does.
Hence, if $T\geq N_{P}$ then $F^T((D_B^i-B^{temp})P_G)$ is computed, otherwise $(F^T(D_B^i-B^{temp}))P_G$ is computed.

The choice of $\mu_1$ and $\mu_2$ has a crucial impact on the rate of
convergence. On the one hand, they should be chosen to get a great
conditioning of the primal problem. On the other hand, these parameters can be
updated during the iterations to improve this rate as it is classically done
in augmented Lagrangian methods~\cite{bertsekas1982constrained}.

A poor conditioning of the primal update appears when these parameters are too
small. The subproblem written in~Eq.~(\ref{eqn:theta_update}) becomes numerically unstable when the elements of the matrix $O(n,t) = D_w(n,n) + D_z(t,t)$ are close to $0$. The eigenvalues of $W = 2\Phi^T \Phi + \mu_1 I_{N_{\Phi}}$ and $Z = \mu_2 P P^T$ are non-negative and their smallest values depend on the penalty parameters.
\\
Let $\{\lambda_w(n), \forall n \in \{1,\cdots,N_{\Phi}\}\}$ and $\{\lambda_z(t), \forall t \in \{1,\cdots,T\}\}$ be respectively the eigenvalues of $2\Phi^T \Phi$ and $PP^T$, we have:
\begin{align}
	\min_{n, t} O(n,t) = \min_{n} \lambda_w(n) + \mu_1 + \mu_2 \min_{t} \lambda_z(t) \enspace .
\end{align}
When $\Phi$ is an overcomplete dictionary, $\min_{n} \lambda_w(n) = 0$
since this is the minimal singular value of $\Phi$. Thus, depending on the
eigenvalues of $PP^T$, $\mu_1$ and $\mu_2$ should be chosen carefully to avoid numerical instability.

An update of these parameters during the iterations can speed up the
convergence. As seen before, the split Bregman scheme is equivalent to the
augmented Lagrangian one when the constraints are linear. Thus, a common
strategy used in the augmented Lagrangian scheme has been chosen here to update
the penalty parameters: when the loss associated with a constraint does not
decrease enough between two iterations, the corresponding parameter is
increased. Formally, let $h_1(X, A) = \|X - A\|_2$ and $h_2(X, B) = \|XP -
B\|_2$ be the constraints losses, the parameters are updated as follows,
\begin{eqnarray}
\label{eqn:updateMu1}
\mu_1^i = \left\{
\begin{array}{ll} 
	\mu_1^{i-1} & \text{if}~~h_1(X^i, A^i) < r_1 h_1(X^{i-1}, A^{i-1})
	\\
	\rho_1 \mu_1^{i-1} & \text{otherwise}
\end{array}
\right.
\end{eqnarray}
and
\begin{eqnarray}
\label{eqn:updateMu2}
\mu_2^i = \left\{
\begin{array}{ll}
	\mu_2^{i-1} & \text{if}~~h_2(X^i, B^i) < r_2 h_2(X^{i-1}, B^{i-1})
	\\
	\rho_2 \mu_2^{i-1} & \text{otherwise}
\end{array}
\right.
\end{eqnarray}
where $r_1$, $r_2$ are the threshold parameters and $\rho_1$, $\rho_2$ are the ratios of the geometric progressions $\mu_1$, $\mu_2$.

To achieve a fast convergence, the initialization of these parameters should not enforce the constraints too strictly in the first iterations but these parameters must not be setup with a value too small in order to not affect the resolution of the primal problem.
A heuristic is described here to ease the initialization of $\mu_1$ and $\mu_2$. This procedure has been shown to be empirically efficient:
\begin{enumerate}
	\item define a search grid $g$ for $\mu_1$ and $\mu_2$,
	\item perform the first iteration of the optimization scheme for each couple of parameters $[g(j), g(l)]$,
	\item evaluate $t_1(j,l) = \frac{\mu_1}{2}h_1(X(j,l)^1, A(j,l)^1)^{2}$ and $t_2(j,l) = \frac{\mu_2}{2}h_2(X(j,l)^1, B(j,l)^1)^{2}$ for each couple,
	\item initialize $\mu_1^0$ and $\mu_2^0$ with:
\begin{align}
	\mu_1^0 ~&=~ g(\argmax_j \sum_{l} t_1(j,l)) \ , \nonumber
	\\ 
	\mu_2^0 ~&=~ g(\argmax_l \sum_{j} t_2(j,l)) \ . \nonumber
\end{align}
\end{enumerate}



\subsection{Full \XX\ algorithm}
\label{subsec:sumup}

\noindent The \XX~is summarized below as a pseudo-code procedure. To ease the understanding of the method, only the crucial steps are highlighted. The initialization and performance optimization steps just point out their corresponding equations. Note that only one subiteration is needed to solve the primal subproblem.

\noindent Parameters: $\lambda_1$, $\lambda_2$, $\mu_1^0$, $\mu_2^0$, $\epsilon$, $iterMax$, $kMax$, $r_1$, $r_2$, $\rho_1$, $\rho_2$

{\small
\begin{algorithmic}
\Procedure{Multi-SSSA}{$Y$, $\Phi$, $P$}
\State Initialize $D_A^0$, $D_B^0$, $X^0$ and set $B^0 = X^0 P$,~ $A^0=X^0$
\State Diagonalize $2\Phi^T \Phi$ and $PP^T$ to get $\Delta_w$, $\Delta_z$, $F$ and $G$
\Comment{Eq.~(\ref{eqn:computeDelta})}
\State Compute $D_w$, $D_z$, from $\Delta_w$, $\Delta_z$, $\mu_1^0$ and $\mu_2^0$)
\Comment{Eq.~(\ref{eqn:computeDiagonal})}
\State Calculate $O$ from $D_w$ and $D_z$
\Comment{Eq.~(\ref{eqn:computeO})}
\State Precompute $Y_{\Phi} = 2F^T \Phi^T Y G$ and $P_G=P^TG$
\State $i = 0$
\While{$i \leq iterMax $ and $\frac{\|X^{i}- X^{i-1}\|_2}{\|X^{i}\|_2} \geq \epsilon $}
\Comment{stopping criteria}
    
	
	\State $M' = Y_{\Phi} - F^T(\mu_1^i (D_A^i-A^{i}) G$ \par 
	\hspace{2.25cm} $ +~~\mu_2^i (D_B^i-B^{i})P_G)$
	\State $X^{i+1} = M' \oslash O$
	\State $X^{i+1} = FX^{i+1} G^T$
	\Comment{$X$ primal update}
	\State $A^{i+1} = \softth_{\frac{\lambda_1}{\mu_1^i}}(X^{i+1} + D_A^i)$
	\Comment{$A$ primal update}
	\State $B^{i+1} = \softth_{\frac{\lambda_2}{\mu_2^i}}(X^{i+1} P + D_B^i)$
	\Comment{$B$ primal update}
	\State $D_A^{i+1} = D_A^i + (X^{i+1} - A^{i+1})$
	\Comment{$D_A$ dual update}
 	\State $D_B^{i+1} = D_B^i + (X^{i+1} P - B^{i+1})$
 	\Comment{$D_B$ dual update}
 	\State Compute $\mu_1^{i+1}$ and $\mu_2^{i+1}$
 	\Comment{Eq.~(\ref{eqn:updateMu1}, \ref{eqn:updateMu2})}
 	\State Update $D_w$, $D_z$ and $O$
 	\Comment{Eq.~(\ref{eqn:computeDiagonal}, \ref{eqn:computeO})}
 	\State $i=i+1$
\EndWhile\\
~~~~\Return $X^{i}$
\EndProcedure
\end{algorithmic}
}


\section{Experiments on synthetic data}

To evaluate the proposed method, two experiments have been performed on synthetic data. The computational time of the algorithm is first evaluated w.r.t. the state-of-the-art. Then, its ability to recover the underlying structure of signals is compared with other classical decomposition approaches. Both experiments have been carried out in the particular case of the fused-LASSO regularization (cf. Eq.~(\ref{eqn:basicProblem}) with $P=P^{\text{TV}}$) on synthetic piecewise constant signals.

\subsection{Data generation}
\label{subsec:data_gen}

\begin{figure}[h!]
	\begin{center}
	\begin{tabular}{lll}
	\includegraphics[scale=0.17]{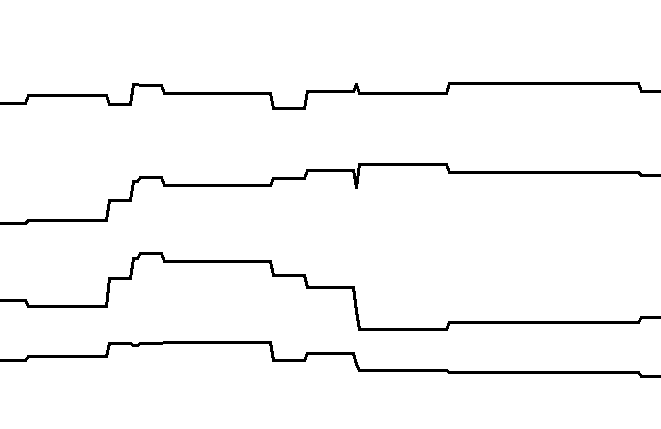}&
	\includegraphics[scale=0.4]{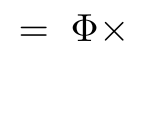}
	\includegraphics[scale=0.17]{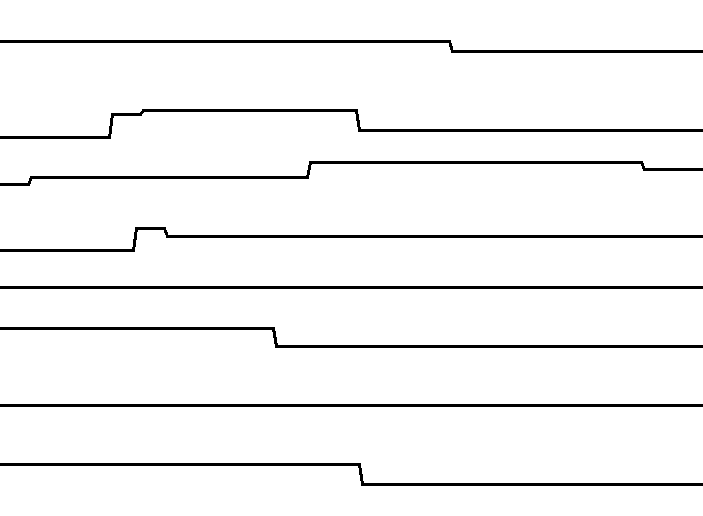}
	\end{tabular}
	\end{center}
	\caption{Example of synthesized signal $Y = \Phi X$, with $C=4$ channels and $N_{\Phi}=8$ atoms.}
	\label{fig:data_synth}
\end{figure}

Each of these piecewise constant signals $Y^k,~~k \in \{1, \cdots, K\}$ has been synthesized from a built decomposition matrix $X^k$ and a dictionary $\Phi$ with $Y^k=\Phi X^k$.
The atoms of $\Phi$ have been drawn independently from a Gaussian distribution to create a random overcomplete dictionary, which consequently has a low coherence:
\begin{align}
\max_{~i, j~\in \{1, \cdots, N_{\Phi}\}, ~i \neq j~} \Phi(i)^T \Phi(j) \approx 0.35 \ . \nonumber
\end{align}
\\
Each block-wise decomposition matrix $X$ has been built as a linear combination of specific activities generated as follows:
\begin{eqnarray}
\Theta_{n, t, d}(i,j) = \left\{
\begin{array}{ll} \nonumber
	0 & \text{if} \ i \neq n
	\\
	{\cal H}(j-(t-\frac{d \times T}{2})) 
	\\
	- {\cal H}(j-(t+\frac{d \times T}{2}))\ \ \ \ & \text{if} \ i = n
\end{array}
\right.
\end{eqnarray}
where $\Theta \in \mathbb{R}^{N_{\Phi} \times T}$, ${\cal H}$ is the Heaviside
function, $n \in \{1, \cdots, N_{\Phi}\}$ the index of an atom, $t$ the
center of the activity and $d$ its duration. Each decomposition matrix $X$
could then be written:
\begin{eqnarray}
	X = \sum_{m=1}^{M} \alpha_m \Theta_{n_m, t_m, d_m} \ , \nonumber
\end{eqnarray}
where $M$ is the number of activities appearing in one signal and the $\alpha_m$ stand for the activation weights.
An example of generated signal is given in Figure~\ref{fig:data_synth}.
 
\subsection{Experimental assessment of computational time}
\label{sec:speed}

The computational time of the \XX\ algorithm is first evaluated. As noticed earlier, the most efficient method proposed for solving the multi-dimensional fused-LASSO is a smooth proximal-gradient method~\cite{chen2012efficient}.
The TV analysis term is approached by a smooth penalty and an accelerated gradient descent is considered for minimizing the cost function. QP and SOCP formulations could also be considered but have been shown to be much slower than the proximal method~\cite{chen2012efficient}. Hence, the proposed optimization scheme is only compared here to this approach on the particular case of the \muu~fused-LASSO.

\subsubsection{Experiments setup}

\paragraph{Tests}
To fairly compare these methods, the unique solution of our convex problem is calculated precisely beforehand. To compute this solution, we used the \XX\ which is stopped when the relative change of the cost between two iterations is under $10^{-10}$. Then, both methods are executed and stopped only when the relative difference between the current cost and the value of the loss at the optimum is under a fixed precision.
Three experiments have been performed to assess the computational time of both methods when different dimensions of our problem are varying. Their respective setups are described in~Figure~\ref{fig:speed_comparison}.
In addition, each test is realized for different values of the precision defined above: $10^{-4}$, $10^{-5}$ and $10^{-6}$.

\begin{figure}
\begin{center}
	\begin{tabular}{|c|c|c|c|}
	\hline
	& $C$ & $N_{\Phi}$ & $T$\\
	\hline
	T1 & 100 & 200 & \begin{tabular}{@{}c@{}} {\scriptsize  $50\etcDots_{50}500$}\\ {\scriptsize $600\etcDots_{100}1000$}\\ {\scriptsize $2000\etcDots_{1000}6000$}\end{tabular}\\
	\hline
	T2 & 100 & \begin{tabular}{@{}c@{}} {\scriptsize $50\etcDots_{50}500$}\\ {\scriptsize $600\etcDots_{100}1000$}\\ {\scriptsize $2000\etcDots_{1000}5000$}\end{tabular} & 300\\
	\hline
	T3 & \begin{tabular}{@{}c@{}} {\scriptsize $50\etcDots_{50}500$}\\ {\scriptsize $600\etcDots_{100}1000$}\\ {\scriptsize $2000\etcDots_{1000}8000$}  \end{tabular} & 200 & 300\\
	\hline
	\end{tabular}
\end{center}
\caption{Settings for speed comparison tests, where $m\etcDots_{n}p = \{m+kn, \forall k \in \mathbb{N}~~s.t.~~m+kn \leq p\}$.}
\label{fig:speed_comparison}
\end{figure}

\paragraph{Regularization parameters} 
The regularization parameter $\lambda_1$ has been fixed such that $\|Y-\Phi \hat{X}\|_2 / \|Y\|_2 \approx 0.1$ where $\hat{X}$ is the output of the optimization method and $\lambda_2$ determined by cross-validation on the distance between the decomposition matrices used to build the signals and those obtained as outputs.

\paragraph{Implementation details}
Both methods have been implemented using MATLAB (64 bits). The experiments have been performed on a PC with 16GB RAM and a 8-core processor.\\
Concerning the proximal approach, the minimization is performed by the classical FISTA method as in \cite{chen2012efficient}. More precisely, the variant named "FISTA with backtracking" in \cite{beck2009fast} has been chosen. The Lipschitzien coefficient $L$ of the smooth term's gradient is approximated then with a variable following a geometrical progression: $L^i = \rho^k L^{i-1}$  with $\rho=1.05$ and $L^0=1$. Besides, the parameter $\mu$ balancing the compromise between the analysis (TV) penalty and its smooth version is chosen such that the precision desired on the solution could be reached. A tight bound of the distance between the strict cost and the smooth one is known theoretically~\cite{chen2012efficient} and is proportional to $\mu$. Consequently, to obtain $finalLoss \times precision \geq gapBound = K_{gap}\mu$ where $finalLoss$ is the minimum value of the cost function and $K_{gap} = \frac{1}{2} N_{\Phi}(T+N_P)$, $\mu$ is defined as follows:
\begin{align}
\mu = 0.95 \times \frac{lossFinal \times precision}{K_{gap}} \ .
\end{align}
As for the introduced scheme, since the initial value of penalty parameters (obtained with the method presented in Section~\ref{sec:implem_penalty}) have been observed to be stable~(for the considered signals), they are computed off-line for each point of our tests on a $20\times20$ logarithmic grid. The update of these parameters is performed with $\rho_1=\rho_2=1.05$ and $r_1 = r_2 = 0.95$. The diagonalizations of $2\Phi^T \Phi$ and $PP^T$ are performed off-line.

\begin{figure}[hbp!]
	\centering
	\subfloat[Test T1]{\includegraphics[trim = 50mm 75mm 50mm 95mm, scale=0.41]{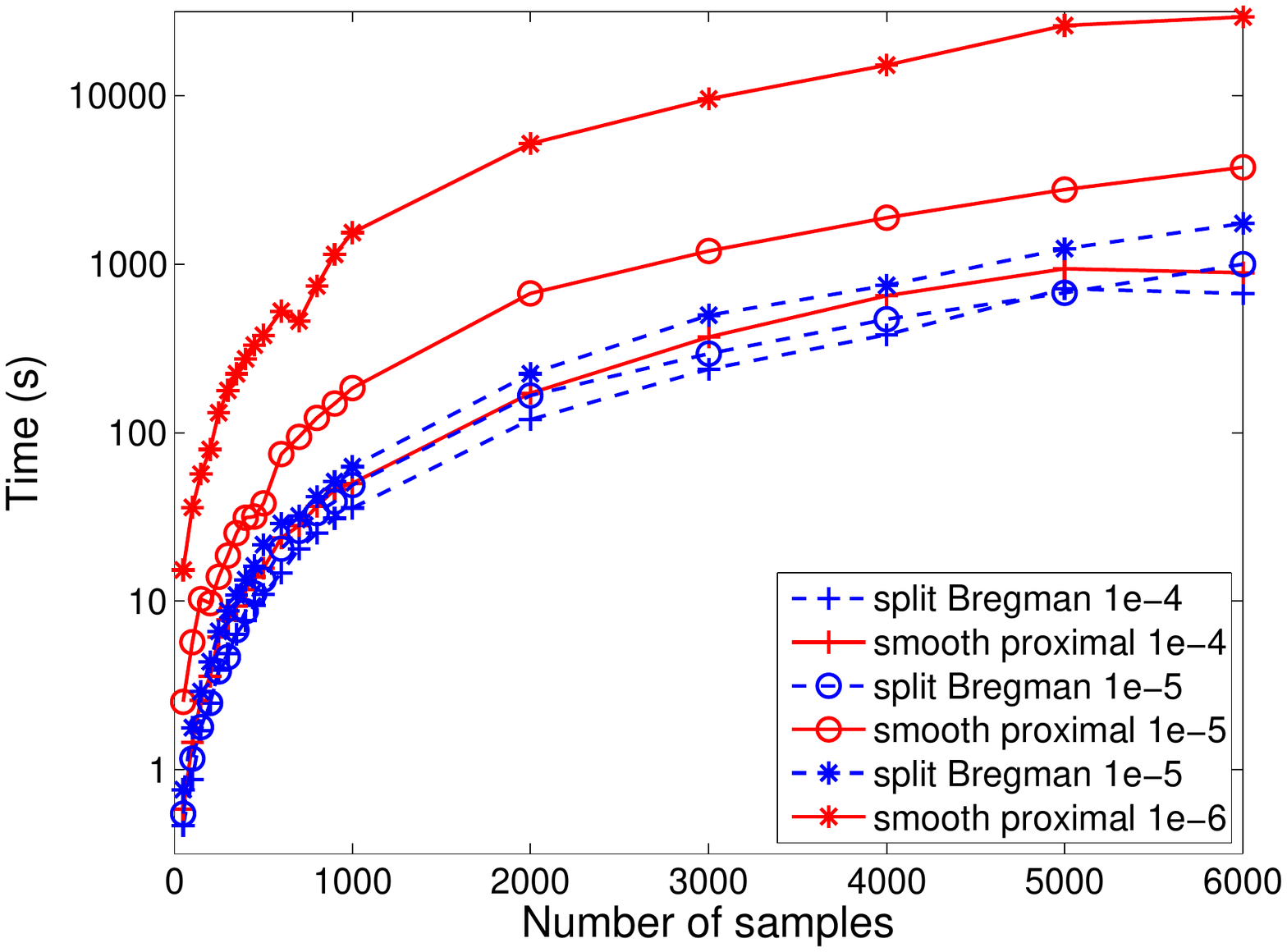}}
	\vspace{0.2cm} \\
	\subfloat[Test T2]{ \includegraphics[trim = 50mm 75mm 50mm 95mm, scale=0.41]{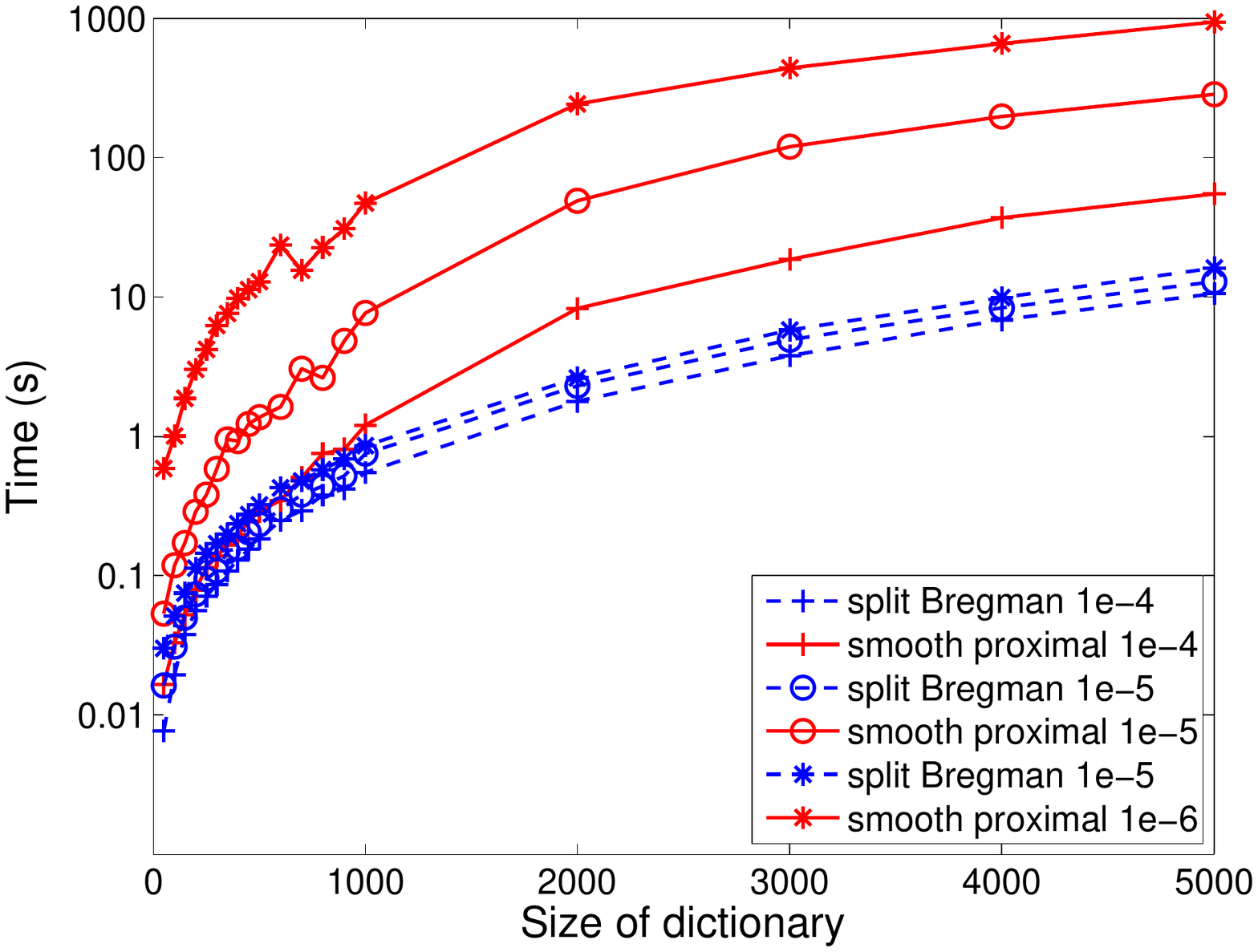}}
	\vspace{0.2cm} \\
	\subfloat[Test T3]{\includegraphics[trim = 50mm 75mm 50mm 95mm, scale=0.41]{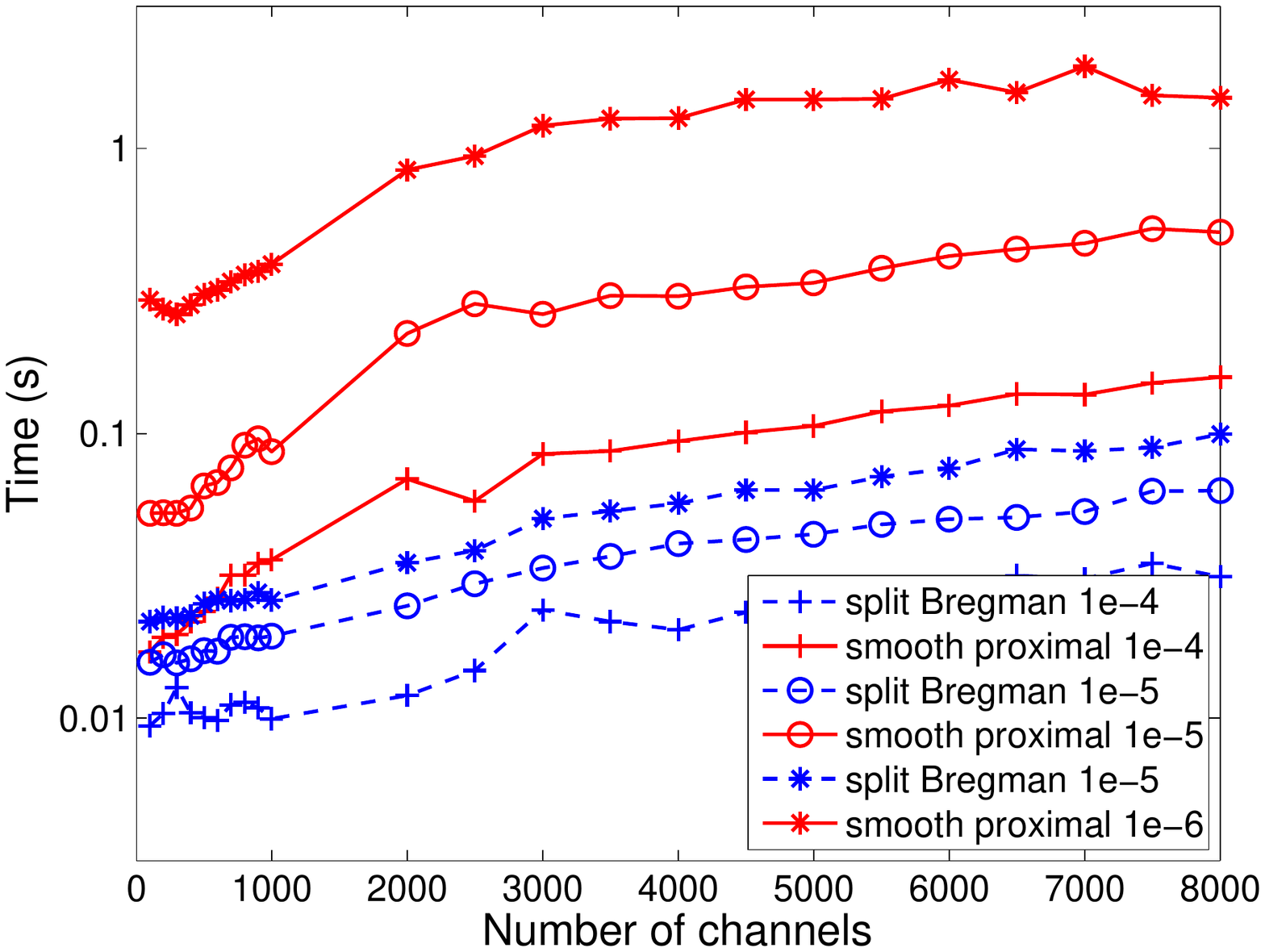}}
	\vspace{0.3cm}
	\caption{Speed comparison: split Bregman (blue) vs smooth proximal gradient (red), for different number of samples (a), size of dictionary (b) and number of channels (c).}
	\label{compVit}
\end{figure}

\subsubsection{Results and discussion}

The results are presented in Figure~\ref{compVit}. The execution times are displayed in a logarithmic scale. As expected, for both methods, the computation time is almost not affected by the number of channels. The observed increase of computation time is due to the computation of the stopping criteria at each iteration. Two more observations can be derived from these curves. Firstly, the split Bregman method is faster than the proximal gradient one in all cases presented here (for all precisions: $10^{-4}$, $10^{-5}$ and $10^{-6}$) and the curves have the same shapes. Secondly, the differences in speed between these approaches become more important when the desired precision becomes smaller. This last observation can be understood by noticing that the gradient's Lipschitzien coefficient of the smooth TV penalty presented in \cite{chen2012efficient} is inversely proportional to the parameter $\mu$. As this coefficient corresponds to the inverse of the FISTA gradient step coefficient, when $\mu$ becomes smaller (insuring a small gap) the gradient step becomes smaller and the method is then slower.


\subsection{Experimental evaluation of sparse recovery on synthetic data}
\label{sec:exp_eval_synth}

The interest of the studied regularization is now studied in a dictionary-based representation context. The performance of the described model to recover the underlying block-wise structures of artificial signals is assessed w.r.t. classical regularizations. 

\subsubsection{Compared regularizations}

\XX\ is compared both with algorithms coding each signal separately with the $\ell_0$ and $\ell_1$ regularization terms and to methods performing the decomposition simultaneously with the $\ell_{2,0}$, $\ell_{2,1}$ and $\ell_{2, 1} + \ell_1$ regularization terms. Regarding the $\ell_0$ and $\ell_{2,0}$ constraints, the solutions are respectively given by the orthogonal matching pursuit (OMP)~\cite{pati1993orthogonal} and the simultaneous OMP (SOMP)~\cite{tropp2006algorithms1}. The $\ell_1$ solutions are obtained by the LARS method~\cite{efron2004least} and a proximal approach (FISTA~\cite{beck2009fast}) has been chosen to deal with the $\ell_{2,1}$ and $\ell_1 + \ell_{2, 1}$ regularizations \cite{gramfort2013time}. The approximation problems with the $\ell_1$ and $\ell_{2,1}$ regularization terms are respectively referred to as the LASSO~\cite{tibshirani1996regression} and group-LASSO~\cite{yuan2006model} problems, where group-LASSO is used defining only one group by atom.

Among the compared algorithms, we have implemented the OMP and the SOMP, whereas the SPAMS\footnote{http://spams-devel.gforge.inria.fr/}~\cite{jenatton2010proximal} toolbox has been used for the other methods.

\subsubsection{Experimental settings}
\label{subsec:exp_setup_synth}

The goal of this experiment is to assess the ability of each model to retrieve the underlying structures of the designed piecewise constant signals. Since these performances vary with the number of activities $M$ composing the signals and their duration $d$, this experience is performed for each point of the following grid of parameters:
\begin{itemize}
 \item $M \in \{20, 30, \cdots, 110\}$ \ ,
 \item $d \sim \mathcal{U}(d_{min}, d_{max})$
 \\
 $(d_{min}, d_{max}) \in \{(0.05, 0.15), (0.15, 0.25), \cdots, (0.95, 1)\}$ \ .
\end{itemize}

For each model and each point in the grid, the evaluation is carried out as follows:
\begin{itemize}
    \item The set of built signals is split to create a training set allowing to determine the best regularization parameters and a test set designed to evaluate the performance with these parameters.
    \item For various regularization parameters, each signal $Y$ of the training set is decomposed and the estimated decomposition matrix $\hat{X}$ is compared with the built one $X$ using the following distance: $\varepsilon (X, \hat{X}) = \|X - \hat{X}\|_{2} / \| X \|_{2}$. The parameters giving the best performances are chosen.
    \item Each signal of the test set is decomposed with the optimal parameters and the same distance is computed between the estimated decomposition matrix and the built one.
\end{itemize}

Other parameters of the signals construction are presented in Figure~\ref{fig:experient_setup}.
\begin{figure}[h!]
\label{fig:experient_setup}
	\begin{center}
    	\begin{tabular}{|ll|ll|}
    		\hline
    		\multicolumn{2}{|c|}{Model} & \multicolumn{2}{|c|}{Activities}
				\\
    		\hline
    		$C=20$&$T=300$ & $t \sim \mathcal{U}(0,T)$ & $\alpha \sim \mathcal{N}(0, 2)$
				\\
    		$N_{\Phi}=30$ & $K=100$ & $n \sim \mathcal{U}(1, N_{\Phi})$ & 
				\\
    		\hline
    	\end{tabular}
    	\caption{Parameters of the experiment of Section~\ref{sec:exp_eval_synth}.}
	\end{center}
\end{figure}

\subsubsection{Results}
\label{subsec:synth_results}

\begin{figure*}[hbp]
	\hfil 
	\centering
	\hspace{-0.4cm}
	\subfloat[Fused-LASSO ($\ell_1+\text{TV}$)]{ \includegraphics[scale=0.21]{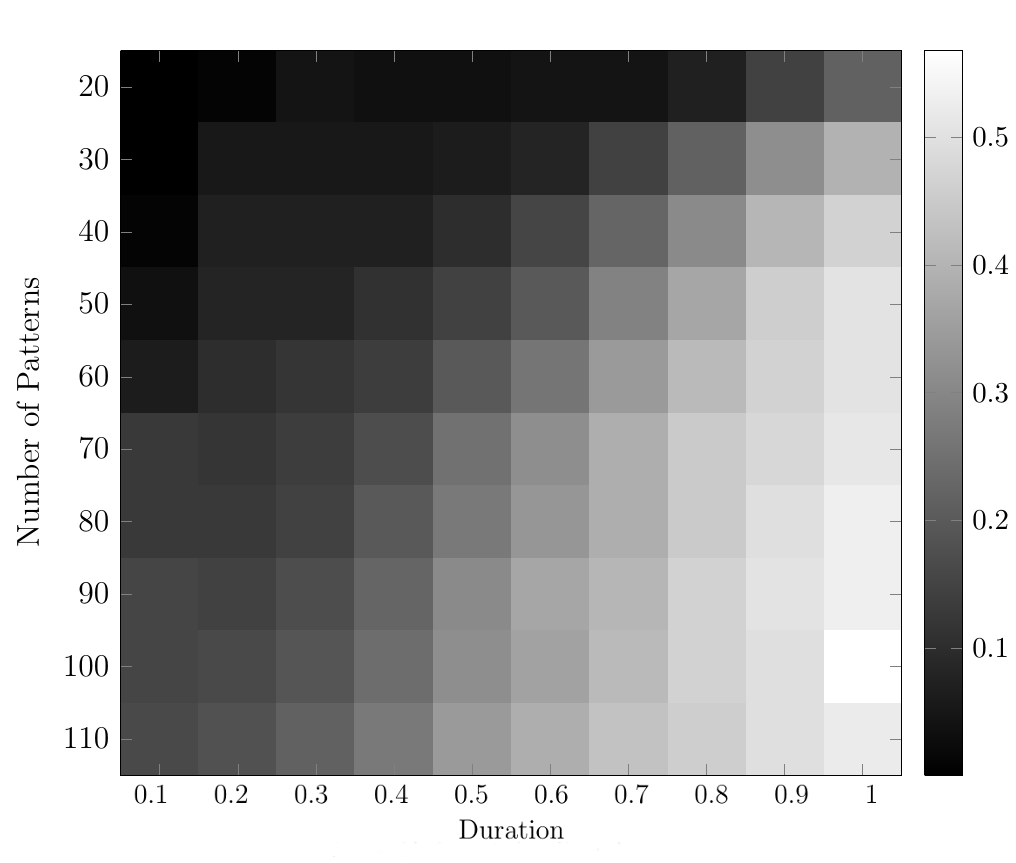} \label{fig:a}}
	\hspace{0.1cm}
	\subfloat[Fused-LASSO vs LASSO ($\ell_1$)]{ \includegraphics[scale=0.21]{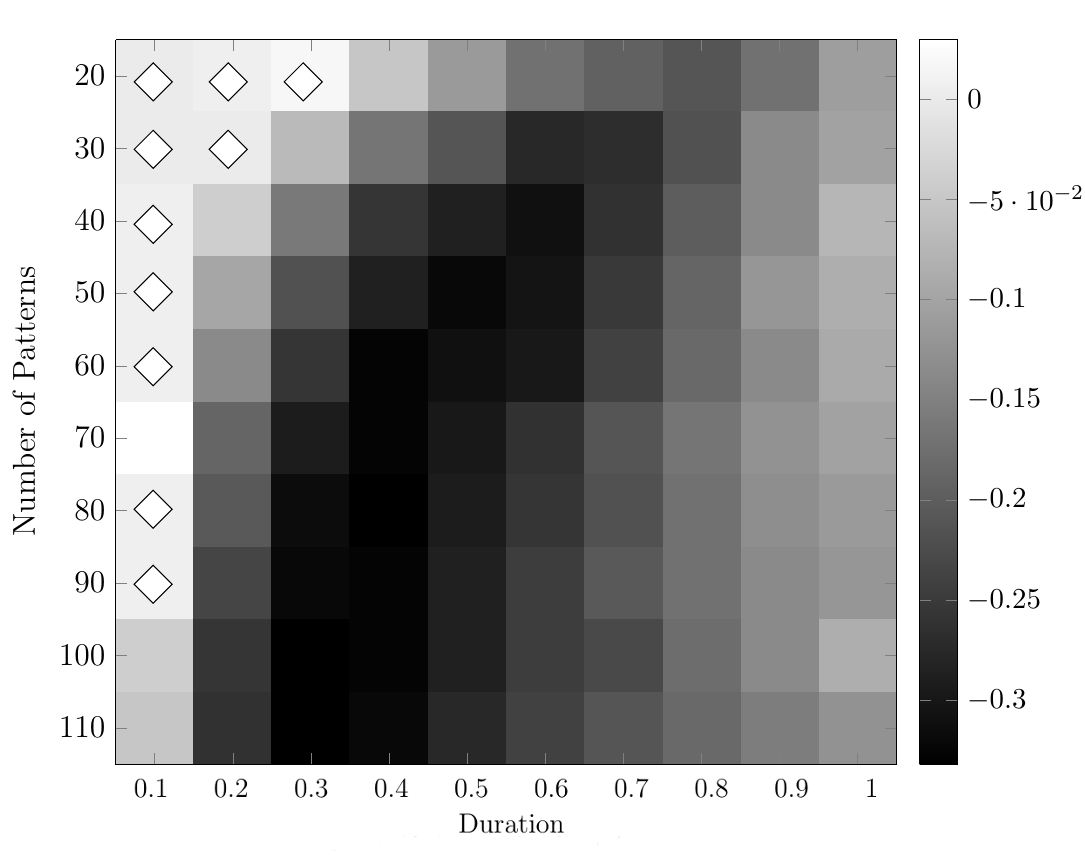} \label{fig:b}}\\
	\vspace{0.3cm}
	\hfil
	\subfloat[Fused-LASSO vs Group-LASSO ($\ell_{2,1}$)]{ \includegraphics[scale=0.20]{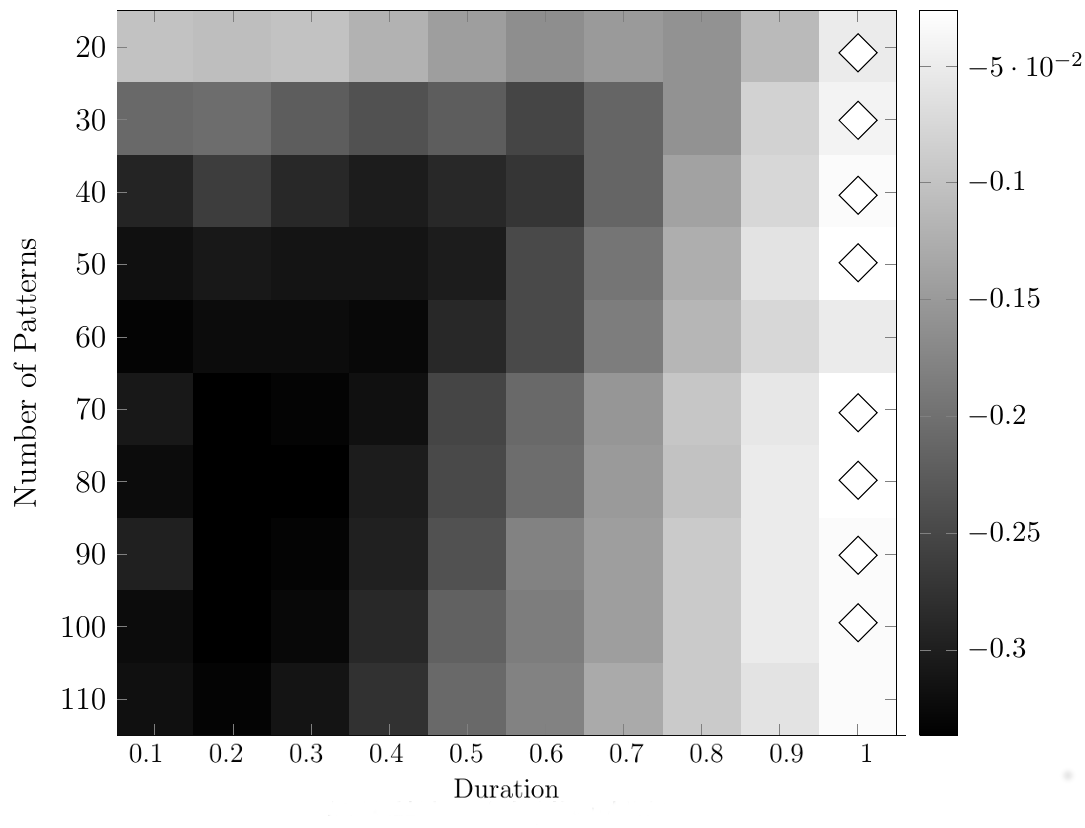} \label{fig:c}}
	\hspace{0.05cm}
	\subfloat[Fused-LASSO vs $\ell_1 + \ell_{2,1}$]{ \includegraphics[scale=0.20]{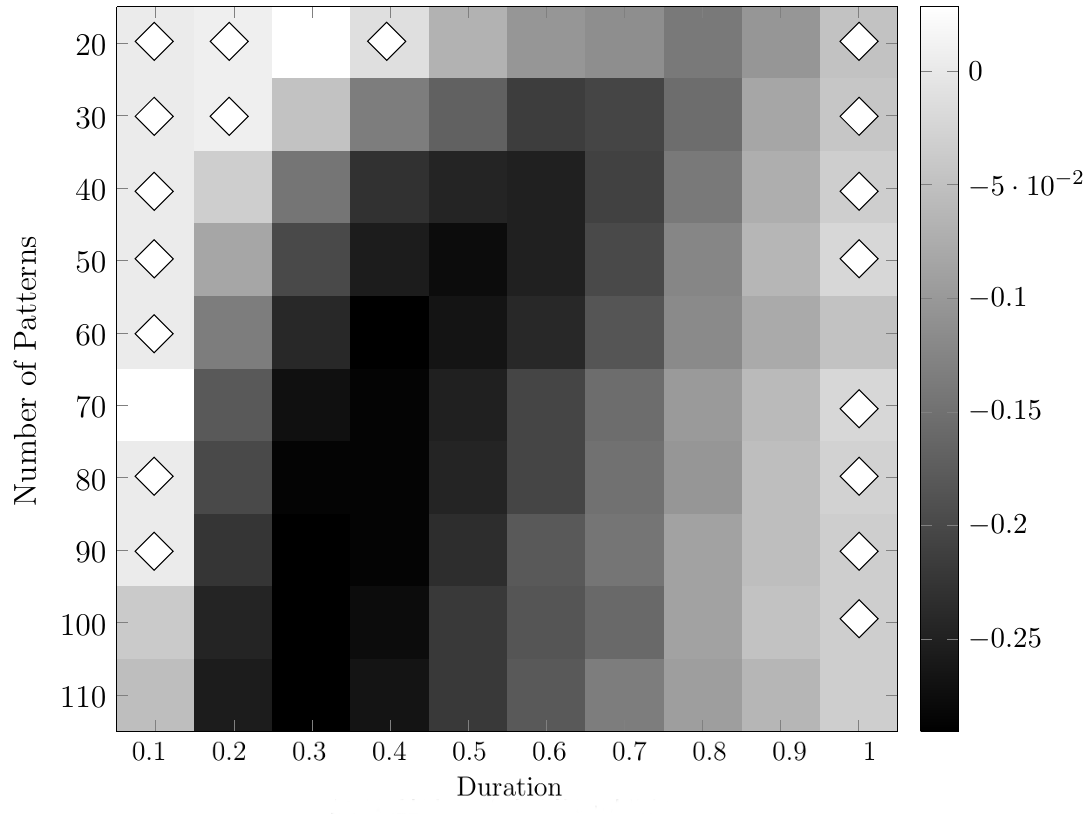} \label{fig:d}}
	\caption{(a) Mean distances obtained with the \XX.~(b) Differences between the
	mean distances obtained with the \XX~and those obtained with the LASSO ($\ell_1$)
	solver.~(c) Differences between the mean distances obtained with the \XX~and
	those obtained with the Group-LASSO ($\ell_{2,1}$) solver.~(d) Differences between the mean
	distances obtained with the \XX~and those obtained with the $\ell_1 + \ell_{2,1}$ solver. The white diamonds correspond to non-significant
	differences.}
	\label{fig:valid}
\end{figure*}

For each point in the grid of parameters, the mean (among test signals) of the distance $\varepsilon$ has been computed for each method and compared to the mean obtained by the \XX~algorithm. First, an analysis of variance shows that significant differences exist between the methods ($p \leq 0.05$). Paired t-tests (with Bonferroni corrections) have been performed to assess the significant differences between the couples of methods.
\\
The results are displayed in Figure~\ref{fig:valid}. In the ordinate axis, the number of patterns $M$ increases from the top to the bottom and in the abscissa axis, the duration $d$ grows from left to right. The top-left image displays the
mean distances obtained by the \XX\ algorithm, \textit{i.e.} $\ell_1+\text{TV}$. Unsurprisingly, the difficulty of finding the true decomposition increases with the number of patterns and their durations. The other figures present its performances compared to other methods by displaying the differences of mean distances in gray scale. These differences are calculated such that negative values (darker blocks) correspond to parameters where the introduced method outperforms the other one. The white diamonds correspond to non-significant differences of mean distances. Results of the OMP and the LASSO solver~($\ell_1$) are very similar as well as those of the SOMP and the group-LASSO solver~($\ell_{2,1}$): they obtain the same pattern of performances on our grid of parameters. So, we only display here the matrices comparing the fused-LASSO regularization to those of the LASSO and group-LASSO models as well as with the regularization $\ell_1 + \ell_{2,1}$.

\subsubsection{Discussions}
\label{subsec:synth_discussion}

First, concerning the comparison between the $\ell_1$ (and $\ell_0$) regularization terms and the $\ell_1+\text{TV}$ one, it can be noted that similar results are obtained when only few atoms are active at the same time. It happens in our artificial signals when only few patterns have been used to create decomposition matrices and/or when the pattern durations are small. On the contrary, when many atoms
are active simultaneously, the \muu~fused-LASSO outperforms the LASSO model, allowing better retrieval of the block-wise structures of signals by using inter-signals prior information.
\\
Concerning the $\ell_{2, 1}$ (and $\ell_{2, 0}$) regularization terms, results depend more on the duration of patterns. When patterns are longer, their performances are similar to the fused-LASSO one. On the contrary, when patterns have short/medium durations the group-LASSO model is outperformed. This is not surprising since these regularization terms select atoms for the entire duration of the signals.
\\
As expected the $\ell_1+\ell_{2, 1}$ model combines the advantages
of the $\ell_1$ and $\ell_{2,1}$ regularization terms, having same performances as
the fused-LASSO for both small number of patterns or long ones. In addition, its efficiency is better than the other studied regularizations in the middle of our grid even if it is still outperformed by the fused-LASSO which better detects the signals abrupt changes.\\
Finally, this experiment illustrates the ability of the method to discriminate activities based on their statistical properties. Interferent signals, which share similar properties with a desired signal, will be split among distinct components if their spatial location of their temporal structure differ.


\section{Application on EEG signals for P300 single-trial classification}
\label{sec:EEG_application}

The general model described in Eq.~(\ref{eqn:basicProblem}) of Section~\ref{sec:prob} can be applied to various contexts, with a non-analytic matrix $P$. This section is devoted to present the application of the studied regularization for the (unsupervised) denoising of real EEG data in a classification task: the detection of P300 evoked potentials.
 
\subsection{Detection of P300 evoked potentials}
The P300 is one of the most popular evoked potential used within Brain Computer Interfaces (BCI) systems. The P300 speller introduced in \cite{farwell1988talking} allows for instance to spell words (letter by letter) by detecting such potentials after the presentation of visual stimuli. These potentials are usually elicited by presenting a rare target stimulus among common non-target ones (oddball approach). The P300 wave appears between 250 and 450 ms after the target stimulus and is mainly located in the parietal and the occipital lobes. Its latency and amplitude depend on various factors like the target-to-target intervals (see \cite{polich2007updating} for a review). The following experiment focuses on the single-trial detection of such brain activity.\\
Each EEG measurement is a spatio-temporal signal which can be studied as a sum of electrical activities emitted by different neural assemblies. These activities are characterized by specific time course and spatial patterns which depend on the locations of their sources and the orientations of the associated electrical activities.\\
In order to obtain plausible decompositions of such signals, the extracted components should respect the properties of these activities as the smoothness induced by the diffusion of electrical waves in the skull. The hypothesis in this experiment is that such plausible representations can be obtained via the studied regularization. The main components of these decompositions could then lead to a better identification of brain activities and in particular the components of the P300 potentials. Two types of noise must be removed: the sensors noise and the EEG background activities not related to the target activity. Since the components of the P300 potential are not known precisely, in order to evaluate the introduced regularization, the main components of the decompositions are used to reconstruct the signals. Then these denoised signals are used to identify P300 potentials.

\subsection{Decomposition model}
\begin{figure}[h!]
	\centering
	\includegraphics[scale=0.35]{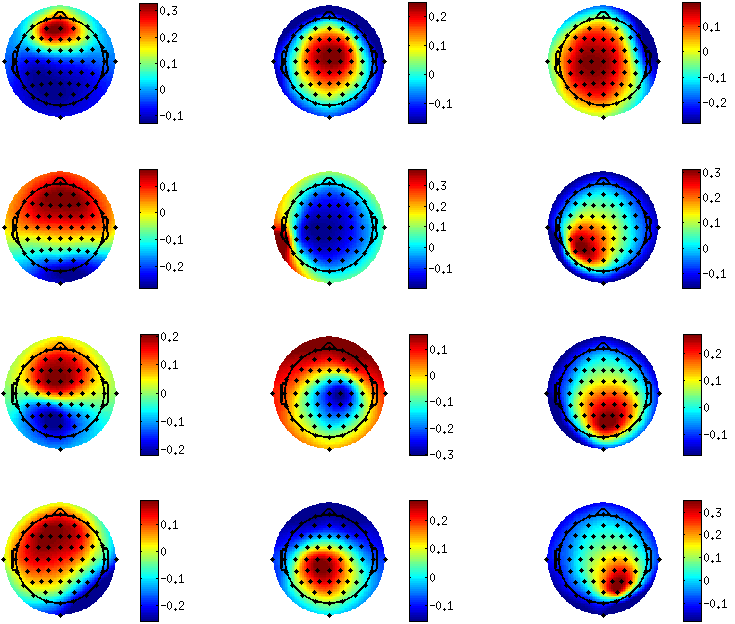}
	\label{fig:spatTopo}
	\caption{Visualization of several topographies of the spatial dictionary $\Phi_{s}$.}
\end{figure}

The previously studied model is considered with EEG signals $Y \in \mathbb{R}^{C \times T}$ recorded on $T$ electrodes over $C$ time samples\footnote{Model in Eq.~(\ref{eqn:basicProblem}) applies a structured prior along the time dimension whereas, in this section, the prior is applied along the spatial dimension. Thus, notations for spatial and temporal dimensions are inversed.}.
These signals are decomposed thanks to the \XX\ algorithm on a Gabor time-frequency dictionary $\Phi \in \mathbb{R}^{C \times N_{\Phi}}$ which is able to efficiently represent both transitory and oscilatory EEG components~\cite{valdes1992frequency}. The analysis regularization matrix $P \in \mathbb{R}^{T \times N_{P}}$ is chosen as the  dual of a spatial dictionary $\Phi_{s} \in \mathbb{R}^{N_{P} \times T}$ composed of $N_{P}$  atoms. This dictionary is composed of realistic EEG topographies, horizontally concatenated since the regularization is applied with the dual on the lines of $X$ (spatial dimension). An EEG topography corresponds to the values of the electrical potential of all electrodes at a given instant. The dual of this dictionary is approximated by the Moore–Penrose pseudo-inverse of $\Phi_{s}$~\cite{elad2007analysis}. The following steps have been carried out to construct this spatial dictionary:
\begin{itemize}
	\item a realistic head model has been built from MRI data and has been divided into voxels,
	\item for each voxel and different orientations of voxels' electrical activities, the associated EEG topographies have been computed by solving the EEG direct problem (with the software OpenMEEG~\cite{gramfort2010openmeeg}) and regrouped in a large spatial dictionary ($\thickapprox$ 5000 elements),
	\item a subset of this large dictionary has been finally selected with a greedy approach such that its coherence does not exceed $0.9$ ($N_{P}=$ 350 elements).
\end{itemize}
Since the retrieval of the sources of these components is not the objective here, this latter step has been carried out to improve the conditioning of the decompositions and preserve a reasonable computational time (decomposition time of a signal around $2$ seconds). Some topographies of $\Phi_s$ are presented in Figure~\ref{fig:spatTopo}.

\subsection{Experimental settings}
\setcounter{paragraph}{0}
In this experiment, the studied regularization (denoted $MSSSA$) is compared to three others: $\ell_1$, $\ell_{2,1}$ and $\ell_{2,1} + \ell_1$ to evaluate their ability to denoise P300 signals and ease their detection.

\paragraph{P300 data}
The dataset IIb of the BCI Competition II \cite{Blankertz2004} has been chosen to evaluate our model. These signals have been recorded on $T=64$ electrodes at 240 Hz in several sessions. Trials have been extracted between 150 and 450 ms after each stimulus and band-filtered between 0.1 and 20 Hz with a fourth order Butterworth filter. The dataset is composed of three sessions and only the two first sessions have been used in this experiment.

\paragraph{Protocol}
To evaluate the efficiency of each regularization, each trial $Y^k$ is firstly decomposed on the dictionary $\Phi$ to obtain its decomposition matrix $\hat{X}^k$, then, a classification algorithm is applied to the set of the  reconstructed signals $\{\hat{Y}^k = \Phi \hat{X}^k, k \in \{1, \dots, K\}\}$. The BLDA (Bayesian Linear Discriminant Analysis) algorithm~\cite{hoffmann2008efficient} has been chosen to perform the classification step. It has been shown particularly efficient for the detection of P300 evoked potentials.\\
The optimal regularization parameters of each model are learned on the second session of the dataset with a $n$-fold cross-validation, and the evaluation is then assessed on the first session of the dataset. This validation is performed for several values of $n$: $2$ ($500$ signals in each fold), $5$ ($200$ signals in each fold) and $10$ ($100$ signals in each fold) in order to evaluate the regularizations efficiency for various sizes of the training set. Each cross-validation is performed $25$ times, with folds selected randomly.\\
For this experiment, the $MSSSA$ has been stopped with the following criterion: $\|X_i-X_{i-1}\|_2 / \|X_i\|_2 \le eps$ (with $eps=10^{-6}$).

\subsection{Results and discussion}

\begin{figure}[h!]
	\centering
	\includegraphics[scale=0.65]{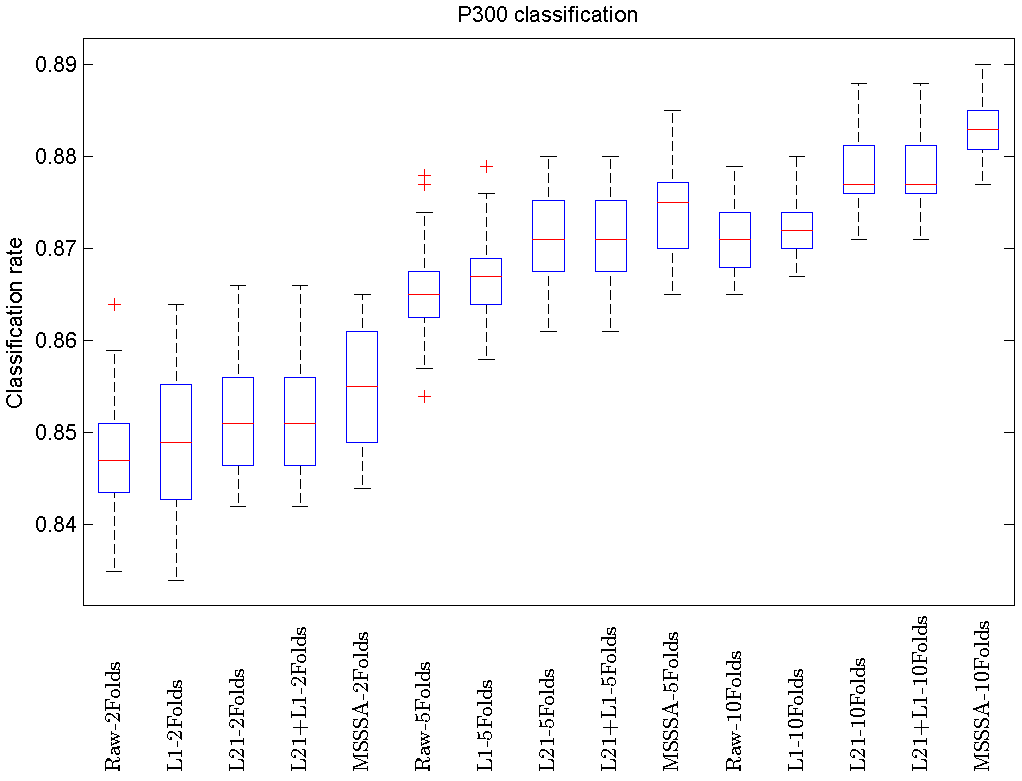}
	\label{fig:resultSimu}
	\caption{Comparison of classification rates between raw signals and rebuilt signals after decomposition with the $\ell_1$($L1$), $\ell_{2,1}$($L21$) and $\ell_{2,1} + \ell_1$($L21+L1$) and \XX\ regularizations (referred to as $MSSSA$), for various number of folds in cross-validation.}
\end{figure}

The results of this experiment are presented in Figure~\ref{fig:resultSimu}. As expected, classification scores rise w.r.t. the number of folds and thus the size of the training set. Concerning the comparison between the regularizations, paired Wilcoxon signed-rank test have been performed between results obtained for each term, given the following results:
\begin{itemize}
\item the $\ell_1$ regularization does not improve the scores obtained on raw signals ($p \ge 0.05$),
\item $\ell_{2,1}$ and $\ell_{2,1} + \ell_1$ regularizations got the same results and improve significantly ($p \le 0.005$) the scores obtained on raw signals. 
\item the introduced $MSSSA$ improve significantly ($p \le 0.005$) the scores obtained with the $\ell_{2,1}$ regularization. 
\end{itemize}

Only the regularizations enforcing a spatial structure in the decomposition coefficients ($\ell_{2,1}$ and $MSSSA$) allow here to improve the classification results by extracting plausible EEG components. The $\ell_{2,1}$ constraints all the channels to be decomposed on the same atoms, enforcing the choice of time-frequency atoms allowing to represent the time courses of all channels. The efficiency of this terms could be explained by noting that brain electrical activities are diffused by the skull and then affect a majority of the sensors. The proposed regularisation improves even more these classification scores by considering a data-driven spatial priors encoded in the analysis matrix $P$ which guide the decomposition to more plausible P300 components.
\\
Removing the noise $E^k = Y^k - \hat{Y}^k$, this decomposition can then be considered as a denoising step before P300 signals classification and more generally before EEG signals classification. In addition, the results could probably be improved by building the spatial dictionary from a head model of the subject on which the EEG are measured.


\section{Conclusion and perspectives}
\label{sec:discussion}
\label{sec:conclu}

The studied approach enforces structural properties of the signal decomposition on the dictionary through the regularization term $\|XP\|_1$. The overall optimization problem is efficiently and scalably handled through the split Bregman method.

Regarding the structural regularization term, a proof of principle of the approach is obtained in the particular case of the TV norm, conducive to the discovery of block-wise structures in the decomposition on overcomplete dictionaries: as shown in Section \ref{sec:exp_eval_synth}, the recovery of the block-wise structure of the input signals
is significantly improved when using the $\ell_1+\text{TV}$ regularization compared to other regularizations.
In addition, the EEG experiment (Section~\ref{sec:EEG_application}) illustrates how the analysis term can be used to enforce structural properties on the decomposition in a data-driven way. It allows in this last case to denoise EEG signals and improve the single-trial detection of P300. The decompositions are guided to plausible components thank to an analysis matrix constructed from a dictionary of realistic EEG topographies.

The second contribution of the paper is the original split-Bregman method 
handling the underlying optimization problem, and its scalability compared with the state-of-the-art. As shown in Section~\ref{sec:speed}, \XX\ outperforms the smooth proximal gradient in terms of speed when the problem dimensions 
increase, and it is less sensitive w.r.t. the wanted precision. Furthermore, 
an empirically efficient heuristic procedure is proposed to adjust the penalty hyper-parameters and thus preserve the efficiency of the algorithm.
\\
This efficiency is explained from the split-Bregman ability to early detect zero coefficients (see \cite{goldstein2009split}, Appendix), thereby easily accommodating $\ell_1$ regularization. The main scalability limitation of the proposed scheme comes from the diagonalization of the matrix in Eq.~(\ref{eqn:computeDelta}), with cubic complexity in its size; it therefore requires the dictionary size $N_\Phi$ and/or $T$ to remain in the hundreds. A second limitation regards the memory complexity, and the storage of variables involved in the method.
\\
To overcome these limitations, a method solving approximatively the primal subproblem Eq.~(\ref{eqn:theta_update}) can be considered. Even if the exact solving of this subproblem allows to ensure the convergence of the proposed scheme, the split Bregman iterations have been shown empirically to converge even when the primal problem is not solved exactly. When $N_{\Phi}$ or $N_{P}$ become really large, this option can significantly reduce the computation time.

Further perspectives and on-going work are primarily concerned with carrying out a full evaluation of the presented EEG application, in particular to examine the influence of the dictionaries choices (spatial and time-frequency) on the denoising efficiency. In addition, an extension of the microstates EEG model will be studied with a TV regularization matrix \cite{Isaac2015a}.
Another direction of research is concerned with approximating the solution of the subproblem in Eq.~(\ref{eqn:impEqn}), instead of solving it exactly.
A mid-term research perspective is to learn both the dictionary $\Phi$ and the regularities $P$ from the data.
\\
While not rigorously shown in this paper, we believe that the proposed framework provides with a rich panel of possibilities to filter signal of interest from noise and interference in various context and applications.




\appendix
\section{Optimization scheme convergence}
\label{appendix1}

The convergence theorem (Section~\ref{subsec:conv}) of the introduced optimization scheme is here demonstrated by applying the convergence analysis of Osher \textit{et al.}~\cite{cai2009split}. This theorem hold for $\lambda_1 \geq 0$, $\lambda_2 \geq 0$, $\mu_1 > 0$ and $\mu_2 > 0$.

The studied iterative algorithm consider at each iteration three convex subproblems Eq.~(\ref{eqn:theta_update}), (\ref{eqn:A_update}), (\ref{eqn:B_update}). The first order optimality condition of these problems gives:
\begin{align}
	\label{eqn:firstSub}
	0 &= (2\Phi^T \Phi + \mu_1 I)X^{i+1} + \mu_2 X^{i+1} P P^T \nonumber
	\\ 
	& ~~ - 2 \Phi^T Y + \mu_1 (D_A^i - A^i) + \mu_2 (D_B^i - B^i)P^T \ , \nonumber 
	\\
	0 &= \lambda_1 Q_A^{i+1} - \mu_1 (D_A^i - A^{i+1} + X^{i+1}) \ , \nonumber
	\\
	0 &= \lambda_2 Q_B^{i+1} - \mu_2 (D_B^i - B^{i+1} + X^{i+1}P) \ ,
	\\
	D_A^{i+1} &= D_A^{i} + (X^{i+1} - A^{i+1}) \ , \nonumber
	\\
	D_B^{i+1} &= D_B^{i} + (X^{i+1}P - B^{i+1}) \ , \nonumber
\end{align}
where $Q_A^{i+1} \in \partial \|A^{i+1}\|_1$ et  $Q_B^{i+1} \in \partial \|B^{i+1}\|_1$.

In addition, the convexity of the main problem Eq.~(\ref{eqn:base}) insures the existence of an unique solution which respect the KKT conditions. The Lagrangian $L$ of the problem could thus be written:
\begin{align*}
	L = \|Y-\Phi X\|_F^2 + \lambda_1 \|X\|_1 + \lambda_2 \|XP\|_1,
\end{align*}
and $\exists \hat{X}$ such that
\begin{align}
	\label{eqn:KKT}
	0 = &- 2 \Phi^T (Y - \Phi \hat{X}) + \lambda_1 \hat{Q}_A + \lambda_2 \hat{Q}_BP^T \ , 
	\\
	&\hat{A} = \hat{X},~~\text{and}~~ \hat{B} = \hat{X}P \ , \nonumber
\end{align}
where $\hat{Q}_A \in \partial \|\hat{A}\|_1$ et $\hat{Q}_B\in \partial \|\hat{B}\|_1$.

This solution is a fixed point of the optimization scheme and verifies:
\begin{align}
	\label{eqn:fixedPoint}
	0 &= (2\Phi^T \Phi + \mu_1 I)\hat{X} + \mu_2 \hat{X} P P^T - 2 \Phi^T Y + \mu_1 (\hat{D}_A - \hat{A}) + \mu_2 (\hat{D}_B - \hat{B})P^T \ , \nonumber 
	\\
	0 &= \lambda_1 \hat{Q}_A - \mu_1 (\hat{D}_A - \hat{A} + \hat{X}) \ , \nonumber
	\\
	0 &= \lambda_2 \hat{Q}_B - \mu_2 (\hat{D}_B - \hat{B} + \hat{X}P) \ ,
	\\
	\hat{D}_A &= \hat{D}_A + (\hat{X} - \hat{A}) \ , \nonumber
	\\
	\hat{D}_B &= \hat{D}_B + (\hat{X}P - \hat{B}) \ . \nonumber
\end{align}

Substracting Eq.~(\ref{eqn:fixedPoint}) from Eq.~(\ref{eqn:firstSub}) we obtained the same system with the errors variables:
\begin{align}
	\tilde{X}^{i} = X^{i} - \hat{X},~\tilde{A}^{i} = A^{i} - \hat{A},~\tilde{B}^{i} = B^{i} - \hat{B} \ , \nonumber
	\\
	\tilde{D}_A^{i} = D_A^{i} - \hat{D}_B,~\tilde{D}_B^{i} = D_B^{i} - \hat{D}_B \nonumber
	\\
	\tilde{Q}_A^{i} = Q_A^{i} - \hat{Q}_A,~\tilde{Q}_B^{i} = Q_B^{i} - \hat{Q}_B \ . \nonumber 
\end{align}

Performing the scalar product of the first line by $\tilde{X}^{i+1}$, the scalar product of the second line by $\tilde{A}^{i+1}$, the scalar product of the third line by $\tilde{B}^{i+1}$ and taking the square Frobenius norm of the last lines, we obtained the following system:
\begin{align*}
	0~ =&~~ 2\|\Phi \tilde{X}^{i+1}\|_F^2 + \mu_1 \|\tilde{X}^{i+1}\|_F^2 + \mu_2 \langle\tilde{X}^{i+1}, \tilde{X}^{i+1}PP^T\rangle
	\\
	&~~ + \mu_1(\langle\tilde{X}^{i+1}, \tilde{D}_A^{i}\rangle - \langle\tilde{X}^{i+1}, \tilde{A}^{i}\rangle) + \mu_2(\langle\tilde{X}^{i+1}, \tilde{D}_B^{i}P^T\rangle - \langle\tilde{X}^{i+1}, \tilde{B}^{i}P^T\rangle) \ ,
	\\
	0~ =&~~\lambda_1 \langle\tilde{A}^{i+1} \tilde{Q}_A^{i+1}\rangle - \mu_1(\langle\tilde{A}^{i+1} \tilde{D}_A^i\rangle - \|\tilde{A}^{i+1}\|_F^2 + \langle\tilde{A}^{i+1}, \tilde{X}^{i+1}\rangle) \ ,
	\\
	0~ =&~~\lambda_2 \langle\tilde{B}^{i+1}, \tilde{Q}_B^{i+1}\rangle - \mu_2(\langle\tilde{B}^{i+1}, \tilde{D}_B^i\rangle - \|\tilde{B}^{i+1}\|_F^2 + \langle\tilde{B}^{i+1}, \tilde{X}^{i+1}P\rangle) \ ,
	\\
	\|\tilde{D}_A^{i+1}\|_F^2 =&~~\|\tilde{D}_A^{i}\|_F^2 + (\|\tilde{X}^{i+1}\|_F^2 + \|\tilde{A}^{i+1}\|_F^2 
	\\
	&~~~-2 \langle\tilde{X}^{i+1}, \tilde{A}^{i+1}\rangle) - 2\langle\tilde{D}_A^{i}, \tilde{X}^{i+1} -\tilde{A}^{i+1}\rangle \ , 
	\\
	\|\tilde{D}_B^{i+1}\|_F^2 =&~~\|\tilde{D}_B^{i}\|_F^2 + (\|\tilde{X}^{i+1}P\|_F^2 + \|\tilde{B}^{i+1}\|_F^2 
	\\
	&~~~-2 \langle\tilde{X}^{i+1}P, \tilde{B}^{i+1}\rangle) - 2\langle\tilde{D}_B^{i}, \tilde{X}^{i+1}P -\tilde{B}^{i+1}\rangle \ .
\end{align*}

Summing the 3 first equations and slightly modifying the others, gives:
\begin{align*}
	0 &=~~2\|\Phi \tilde{X}^{i+1}\|_F^2 + \mu_1 \|\tilde{X}^{i+1}\|_F^2 + \mu_2 \langle\tilde{X}^{i+1}, \tilde{X}^{i+1}PP^T\rangle 
	\\
	&~~+ \lambda_1 \langle\tilde{A}^{i+1}, \tilde{Q}_A^{i+1}\rangle +  \lambda_2 \langle\tilde{B}^{i+1} , \tilde{Q}_B^{i+1}\rangle 
	\\
	&~~+ \mu_1(\langle\tilde{X}^{i+1}, \tilde{D}_A^{i}\rangle - \langle\tilde{X}^{i+1}, \tilde{A}^{i}\rangle - \langle\tilde{A}^{i+1}, \tilde{D}_A^i\rangle +\|\tilde{A}^{i+1}\|_F^2 - \langle\tilde{A}^{i+1}, \tilde{X}^{i+1}\rangle)
	\\
	&~~+\mu_2(\langle\tilde{X}^{i+1}, \tilde{D}_B^{i}P^T\rangle - \langle\tilde{X}^{i+1}, \tilde{B}^{i}P^T\rangle -\langle\tilde{B}^{i+1}, \tilde{D}_B^i\rangle + \|\tilde{B}^{i+1}\|_F^2 - \langle\tilde{B}^{i+1}, \tilde{X}^{i+1}P\rangle) \ ,
	\\
	\langle\tilde{D}_A^{i},& \tilde{X}^{i+1} -\tilde{A}^{i+1}\rangle =~~\frac{1}{2}(\|\tilde{D}_A^{i+1}\|_F^2 - \|\tilde{D}_A^{i}\|_F^2 - \|\tilde{X}^{i+1} - \tilde{A}^{i+1}\|_F^2) \ ,
	\\
	\langle\tilde{D}_B^{i},& \tilde{X}^{i+1}P -\tilde{B}^{i+1}\rangle =~~\frac{1}{2}(\|\tilde{D}_B^{i+1}\|_F^2 - \|\tilde{D}_B^{i}\|_F^2 - \|\tilde{X}^{i+1}P - \tilde{B}^{i+1}\|_F^2) \ .
\end{align*}
and combining these equations and summing between $i=1$ and  $i=I$:
\begin{align*}
	&\frac{\mu_1}{2}(\|\tilde{D}_A^{1}\|_F^2 - \|\tilde{D}_A^{S}\|_F^2) + \frac{\mu_2}{2}(\|\tilde{D}_B^{1}\|_F^2 - \|\tilde{D}_B^{S}\|_F^2)
	\\
	&~~~=~2 \sum_{i=1}^{S}\|\Phi \tilde{X}^{i}\|_F^2 + \sum_{i=1}^{S} \lambda_1 \langle\tilde{A}^{i+1}, \tilde{Q}_A^{i+1}\rangle +  \lambda_2 \langle\tilde{B}^{i+1} , \tilde{Q}_B^{i+1}\rangle
	\\
	&~~~+ \frac{\mu_1}{2}(- \|\tilde{A}^{1}\|_F^2 + \sum_{i=1}^{S}\|\tilde{X}^{i+1} - \tilde{A}^{i+1}\|_F^2 + \|\tilde{X}^{i+1} - \tilde{A}^{i}\|_F^2 +\|A^{S}\|_F^2)
	\\
	&~~~+ \frac{\mu_2}{2}(- \|\tilde{B}^{1}\|_F^2 + \sum_{i=1}^{S}\|\tilde{X}^{i+1}P - \tilde{B}^{i+1}\|_F^2 + \|\tilde{X}^{i+1}P - \tilde{B}^{i}\|_F^2 + \|B^{S}\|_F^2) \ .
\end{align*}

The convexity of $\|.\|_1$ imply that the terms $<\tilde{A}^{i}, \tilde{Q_A}^{i}>$ and $<\tilde{B}^{i}, \tilde{Q_B}^{i}>$ are positives ($\forall i$). $\mu_1$, $\mu_2$, $\lambda_1$ and $\lambda_2$ being non-negative, all terms of the above equations are non-negative and we have:
\begin{align*}
	& \frac{\mu_1}{2}(\|\tilde{D}_A^{1}\|_F^2 +\|\tilde{A}^{1}\|_F^2) + \frac{\mu_2}{2}(\|\tilde{D}_B^{1}\|_F^2 + \|\tilde{B}^{1}\|_F^2)
	\\
	&~~~~\ge  ~2\sum_{i=1}^{S}\|\Phi \tilde{X}^{i}\|_F^2 + \sum_{i=1}^{S} \lambda_1 \langle\tilde{A}^{i+1}, \tilde{Q}_A^{i+1}\rangle + \lambda_2 \langle\tilde{B}^{i+1} , \tilde{Q}_B^{i+1}\rangle 
	\\
	&~~~~+ \frac{\mu_1}{2} \|\tilde{X}^{i+1} - \tilde{A}^{i}\|_F^2 \nonumber + \frac{\mu_2}{2} \|\tilde{X}^{i+1}P - \tilde{B}^{i}\|_F^2 \ .
\end{align*}
\noindent From this last equation we can derive:
\begin{align}
	\label{eqn:firstR}
	\sum_{i=1}^{\infty}\|\Phi \tilde{X}^{i}\|_F^2 < \infty \ ,
	\\
	\label{eqn:secondR}
	\sum_{i=1}^{\infty} \langle\tilde{A}^{i+1}, \tilde{Q}_A^{i+1}\rangle~ < \infty,~\sum_{i=1}^{\infty} \langle\tilde{B}^{i+1} , \tilde{Q}_B^{i+1}\rangle~ < \infty \ , 
	\\
	\label{eqn:thirdR}
	\sum_{i=1}^{\infty} \|\tilde{X}^{i+1} - \tilde{A}^{i}\|_F^2~ < \infty,~\sum_{i=1}^{\infty} \|\tilde{X}^{i+1}P - \tilde{B}^{i}\|_F^2~ < \infty \ ,
\end{align}
which leads to the convergence theorem enunciated in Section~\ref{eqn:th_}.

\noindent From Eq.~(\ref{eqn:secondR}) and the properties of the Bregman distance (cf. Eq.~(3.6) in \cite{cai2009split}):
\begin{align}
	\lim_{i  \to \infty} \|A^i\|_1 - \|\hat{A}\|_1 - \langle A^i - \hat{A},~ \hat{Q}_A\rangle = 0 \ , \nonumber
	\\
	\lim_{i  \to \infty} \|B^i\|_1 - \|\hat{B}\|_1 - \langle B^i - \hat{B},~ \hat{Q}_B\rangle = 0 \ , \nonumber
\end{align}
which, combined with Eq.~(\ref{eqn:thirdR}), gives:
\begin{align}
	\label{eqn:firstF}
	\lim_{i  \to \infty} \|X^i\|_1 - \|\hat{X}\|_1 - \langle X^i - \hat{X},~ \hat{Q}_A\rangle = 0 \ ,
	\\
	\label{eqn:secondF}
	\lim_{i  \to \infty} \|X^iP\|_1 - \|\hat{X}P\|_1 - \langle X^i - \hat{X},~ \hat{Q}_B P^T\rangle = 0 \ ,
\end{align}
and finally, by taking $\lambda_1$ Eq.~(\ref{eqn:firstF}) + $\lambda_2$ Eq.~(\ref{eqn:secondF}) and Eq.~(\ref{eqn:KKT}), we have:
\begin{align}
	\label{eqn:last}
	\lim_{i \to \infty} &\|X^i\|_1 - \|\hat{X}\|_1 + \|X^iP\|_1 - \|\hat{X}P\|_1 \nonumber
	\\
	&- \langle X^i - \hat{X},~ 2\Phi^T(Y-\Phi \hat{X})\rangle = 0 \ .
\end{align} 

In addition, $\|\Phi \tilde{X}^i\|_2^2 = <\nabla f(X^i) - \nabla f(\hat{X}), X^i -\hat{X}> $ for $f(X) = \|Y - \Phi X\|_2^2$. $F$ is convex and from Eq.~(\ref{eqn:firstR}) and the same Bregman distance's property used before we can write:
\begin{align}
	\lim_{i \to \infty} &\|Y - \Phi X^i\|_F^2 - \|Y - \Phi \hat{X}\|_F^2 \nonumber
	\\
	&- \langle X^i - \hat{X}, -2 \Phi^T (Y-\Phi \hat{X})\rangle \ , \nonumber
\end{align}
which provide with Eq.~(\ref{eqn:last}) the first result of the theorem:
\begin{align}
	\lim_{i \to \infty}  &\|Y - \Phi X^i\|_F^2 + \lambda_1 \|X^i\|_1 + \lambda_2 \|X^i P\|_1 \nonumber 
	\\
	= &\|Y - \Phi \hat{X}\|_F^2 + \lambda_1 \|\hat{X}\|_1 + \lambda_2 \|\hat{X} P\|_1 = 0 \ . \nonumber
\end{align}

The second part is obtained by noting that the fonction $g(X) = \|Y - \Phi X\|_2^2 + \lambda_1 \|X\|_1 + \lambda_2 \|X P\|_1$ is continuous, strictly convex and then has an unique minimizer. So, $\lim_{i \to \infty} g(X^i) = g(\hat{X}) \Rightarrow \lim_{i \to \infty}  X^i = \hat{X}$ (see Osher \textit{et al.}~\cite{cai2009split}).


\bibliographystyle{elsarticle-harv}
\bibliography{biblio}

\end{document}